\documentclass[onecolumn]{emulateapj}
\usepackage{graphicx}
\usepackage{amssymb}
\usepackage{array}

\shorttitle{LAUNCHING AND QUENCHING OF BLACK HOLE RELATIVISTIC JET}
\shortauthors{Pu, Hirotani, \& Chang}

\begin{document}

\title{Launching and Quenching of Black Hole Relativistic Jets \\
    At Low Accretion Rate}

\author{Hung-Yi Pu\altaffilmark{1}, Kouichi Hirotani\altaffilmark{2}, and Hsiang-Kuang Chang\altaffilmark{1,3}}

\altaffiltext{1}{Department of Physics, National Tsing Hua University, Hsinchu
30013, Taiwan}
\altaffiltext{2}{Theoretical Institute for Advanced Research in Astrophysics,
Academia Sinica, Institute of Astronomy and Astrophysics, P.O. Box
23-141, Taipei, Taiwan}
\altaffiltext{3}{Institute of Astronomy, National Tsing Hua University, Hsinchu
30013, Taiwan}

\begin{abstract}
Relativistic jets are launched from black hole (BH) X-ray binaries and active galactic nuclei  when the disk accretion rate is below a certain limit (i.e., when the ratio of the accretion rate to the Eddingtion accretion rate, $\dot{m}$, is below about $0.01$) but quenched when above. We propose a new paradigm to explain this observed coupling between the jet and the accretion disk by investigating the extraction of the rotational energy of a BH when it is surrounded by different types of accretion disk. At low accretion rates (e.g., when $\dot{m}\lesssim0.1$), the accretion near the event horizon is quasi-spherical. The accreting plasmas fall onto the event horizon in a wide range of latitudes, breaking down the force-free approximation near the horizon. To incorporate the plasma inertia effect, we consider the magnetohydrodynamical (MHD) extraction of the rotational energy from BHs
by  the accreting MHD fluid, as described by the MHD Penrose process. It is found that the energy
extraction operates, and hence a relativistic jet is launched, preferentially when the accretion disk consists of an outer Shakura-Sunyaev disk (SSD) and an inner advection-dominated accretion flow. When the entire accretion disk type changes into an SSD, the jet is quenched because the plasmas brings more rest-mass energy than what is extracted from the hole electromagnetically to stop the extraction.
Several other observed BH disk-jet couplings, such as why the radio 
luminosity increases with increasing X-ray luminosity until the radio emission drops, are also explained.
\end{abstract}


\section{Introduction}
Black hole (BH) relativistic jets  are
observed from black hole X-ray binaries (BHXBs) and active galactic
nuclei(AGNs).
A puzzling coupling between the jet and the
accretion disk has been recognized.
Observationally,
the radio luminosity
(which is likely due to the radiation from the relativistic jet or may also be
from the semi-relativistic disk wind) increases with increasing X-ray
luminosity (which is likely due to the radiation from the disk), until it shows
a sudden drop when the X-ray luminosity exceeds a certain limit \citep{gal03,mac03,fen04,tru11}.
In terms of the dimensionless accretion rate $\dot{m}$,
this limit has a value about $\sim0.01$, where $\dot{m}\equiv\dot{M}/\dot{M}_{\mathrm{Edd}}$,
$\dot{M}$ is the accretion rate, $\dot{M}_{\mathrm{Edd}}=L_{\mathrm{Edd}}/\chi c^{2}$ is
the Eddingtion accretion rate, $L_{\mathrm{Edd}}=1.38\times10^{38}(M/M_{\odot})\mathrm{erg\, s^{-1}}$
is the Eddingtion luminosity, $\chi$ is the efficiency converting the energy of the accreting
mass into the radiation energy, and $c$ is the speed of light.
In addition, when the X-ray luminosity changes, the X-ray
spectral state also changes, indicating that the accretion disk
changes its type with the accretion rate \citep{fen04,tru11,esi97,rem06}.
Since it is believed that BH relativistic jets are originated from the rotating BHs
that are threaded by large-scale magnetic lines and are powered by
the rotational energy of the BH \citep{bla77},
an investigation of  whether the BH energy can be 
extracted by hole-threading magnetic field lines when the BH
is surrounded by different type of accretion disks,
may offers the key to an understanding of the disk-jet coupling of BHXBs
and AGNs.

%although the accretion disks
%are adequate for ejecting semi-relativistic wind \citep{bla82,sad10}.

How the disk type varies $\dot{m}$ is described below.
When the accretion rate is extremely low, the energy transfer due
to Coulomb collisions between ions and electrons is very inefficient.
As a result, the ions cannot efficiently transfer the heat to the
electrons, and the disk becomes an advection-dominated accretion flow
or ADAF for short \citep{nar95,nar97,nar98}, which corresponds to a hot, radiatively inefficient, optically thin 
and geometrically thick disk. As the accretion rate increases,
the outer part of the disk cools down due to efficient Coulomb collisions,
and becomes a cold, optically thick and geometrically thin disk \citep{sha73,nov73},
hereafter
we use the term "SSD " (Shakura-Sunyaev disk) to refer to such disk; 
consequently, a combined
disk is formed with an inner ADAF and an outer SSD.
In a combined disk, the radius where the transition from thin disk to ADAF
occurs is called the transition radius, $R_{\mathrm{tr}}$. In general,
$R_{\mathrm{tr}}$ decreases with increasing $\dot{m}$ \citep{hon96,man00}.
When the accretion
rate further increases and exceeds a critical value with $\dot{m}\sim0.01$,
 the Coulomb interaction
becomes so efficient at every radius of the disk that the whole disk
cools down, changing into the SSD. The plasma rotational motion is
nearly Keplerian in a SSD, whereas it becomes sub-Keplerian in an
ADAF, because the pressure gradient force contributed in the radial
force balance in the latter type of accretion disk.

It is interesting to note that the accretion flow geometries \textit{near the horizon} are 
\textit{quasi-spherical} for all the above disks (see Section 2 for more details). As 
a result, there are considerable plasma flow along the hole-threading line,
reducing the electromagnetic extraction of BH energy by contributing
their rest-mass energy. This consideration mainly differentiates our model from previous 
explanations of the disk-jet couplings (e.g., Meier 2005; Ferreira et al. 2006).
We emphasize that this effect is important at low accretion rate, e.g., when $\dot{m}\lesssim0.1$, 
due to the quasi-spherical geometry near the event horizon;
at higher accretion rate
, e.g., when $\dot{m}$ as large as $\gtrsim0.3$ \citep{abr10}, the accretion disk
type further varies to a slim disk \citep{abr88,abr10,sad09} and a \textit{disk-type} geometry near the horizon
is realized. Therefore, a pure electromagnetic extraction of BH energy at most of the latitudes
becomes a good approximation.

To consider the contribution of the quasi-spherical accreting plasma near the BH, we adopt the magnetohydrodynamics (MHD) theory and  consider the {}``MHD
 Penrose process'' \citep{tak90}, which includes both the electromagnetic
and plasma inertia effects, to examine the extraction of the BH rotational energy\footnote{
In comparison, in the magnetically dominated limit, the Blandford-Znajek process \citep{bla77} becomes a good approximation of the MHD Penrose process.}.
By constraining the magnetic field strength and the plasma density from the theory of accretion disks in a strong gravitational field, we investigate the importance of plasma inertia for different accretion disk types at low accretion rates. It is found that the MHD extraction of BH energy takes place preferentially in a combined disk and therefore enables the launch of a relativistic jet; however, such relativistic jets are quenched if a magnetic dominance breaks down when the entire accretion disk become a thin disk. We also propose a new paradigm of the coupling between the accretion disk and the jet for both BHXBs and AGNs. The observed launching and quenching of relativistic jets at low accretion rates and several other puzzling observational results, are naturally explained.

We introduce our model in Section 2. In Section 3, 
we summarize necessary formulae and concepts of the general
relativistic ideal MHD flow in a stationary
and axisymmetric BH magnetosphere. 
Next, the result is given in Section 4. Finally, discussion and
conclusion are respectively presented in Section 5 and 6.

\section{The Model}

To qualitatively investigate whether a relativistic jet can be launched when the BH is 
surrounded by different type of accretion disks, we consider a model
with  following assumptions. Comments on the assumptions are provided in Section 5.3.

\textit{1. Accretion disks are described by the analytical solutions adopted from the accretion disk theories.}
The accretion disk type is important since 
both the large-scale field strength and the mass flux per magnetic flux tube
(hereafter, mass loading) are related to the disk. 
For simplicity, we adopt
the analytical solutions of the ADAF and the SSD
to represent the disk properties. For a combined disk, 
these two types of disk solutions are connected at the transition radius.
The analytical solution we adopted is provided in the Appendix B. The $\lq$middle region' solution of a SSD, in which the pressure and the
opacity is respectively dominated by the gas pressure and the electron-scattering opacity, is used.

\textit{2. Parabolic large-scale hole-threading magnetic field lines are considered.}
We consider large-scale magnetic field lines that are dragged in from
a distant region by the accretion flow. It is expected that 
the field lines can be finally accumulated near the
BH with a parabolic geometry. Thus, instead of solving
for the field geometry, we  adopt the paraboloidal
field solution given in \citet{bla77} for the magnetic flux function
$A_{\phi}$, \begin{equation}
A_{\phi}=\frac{\mathbb{\mathcal{H}}}{2}\left\{ r(1-\cos\theta)+2\left(1+\cos\theta\right)\left[1-\ln\left(1+\cos\theta\right)\right]\right\} ,\end{equation}
where the field strength $\mathcal{H}$ is to be determined (see Appendix C for more details). 
Here, we focus on magnetic field lines that thread the horizon, since they are responsible
for the extraction of the rotational energy from a rotating BH.

\textit{3. The accretion flow has a quasi-spherical geometry near the event horizon.}
An accretion flow will have a quasi-spherical geometry
near the horizon
if the flow become transonic before the radius of the 
innermost stable circular orbit, $r_{\rm{ISCO}}$,  \citep{abr81}. Compare to a disk-like geometry,
in which the accreting plasma enters the BH through a plunging region 
near the equatorial plane, the accreting plasma in a quasi-spherical geometry 
can fall onto 
the horizon at a wider range of latitudes.
Since the sonic point of the accretion flow is well outside of  $r_{\rm{ISCO}}$
for a typical ADAF,  which has $\alpha\gtrsim0.2$ \citep{nar97},
and the sonic point is close to $r_{\rm{ISCO}}$  for an SSD,
the plasma distribution \textit{near the horizon} should have a quasi-spherical
geometry for all the disk types we considered here, ADAF, combined disk, and SSD.
Note that, for the SSD, the plasmas are conventionally 
supposed  to plunge into the horizon in a disk-like geometry, 
which forms a striking contrast with our current picture.

\textit{4. The magnetic field strength is parameterized by the gravitational binding energy.}
%\noindent
Providing that the diffuse effect of the large-scale magnetic 
field is relatively unimportant, large-scale magnetic field lines
can be {}`arrested' by the accretion disk, being dragged from the
outer region toward the central BH \citep{nar03,spr05,rot08}.  It is convenient
to parameterize its strength $B$ at radius $R$ by (see Appendix A)
\begin{equation}
B\equiv B(\varepsilon,R,\sigma(R))=\varepsilon\sqrt{2\pi GM_{\mathrm{BH}}\sigma/R^{2}}\;,
\end{equation}
where $G$ is the gravitational constant, $M_{\mathrm{BH}}$ is the mass of the central
BH , $\sigma$ is the surface density of the disk, and $\varepsilon^{2}$
is the ratio of the gravitational binding energy of the disk at $R$
to the large-scale magnetic field energy inside radius $R$.
The determination of $\varepsilon$ is not yet clear; however, it
can be related to the ionization degree of the accretion flow and
the magnetic Prandtl number. 

\textit{5. The accretion disk type is a function of the accretion rate.}
As described in Section 1, from low to high $\dot{m}$, the disk type varies as follows: 
(I) ADAF, (II) combined disk that consists
of an inner ADAF and outer SSD, and
(III) SSD. Hereafter, we call the corresponding range of $\dot{m}$ as {}``Range
I'', {}``Range II'', and {}``Range III'', respectively.
Many studies on the transition radius $R_{\mathrm{tr}}$ between inner
ADAF and outer SSD give similar results \citep{hon96,man00}.
To qualitatively describe how the disk configuration changes
with $\dot{m}$ , we adopt the analytical solution of $R_{\mathrm{tr}}$ from \citet{hon96}
\begin{equation}
R_{\mathrm{tr}}=2.7648\alpha^{4}\dot{m}{}^{-2}R_{\mathrm{\mathrm{g}}}\qquad\qquad\mathrm{if}\:\dot{m}\leq\dot{m}_{\mathrm{cr}},
\end{equation}
where $\alpha$ is
the viscosity of ADAF in the $\alpha$-prescription \citep{sha73}, $\dot{m}_{\mathrm{cr}}=\dot{M}_{\mathrm{cr}}/\dot{M}_{\mathrm{Edd}}$
is the critical accretion rate with a typical value $\leq0.01$, beyond
which there is no ADAF solution and hence the entire disk becomes
the SSD type. 
We recognize that the entire disk becomes thin disk if $R_{\mathrm{tr}}<10R_{\mathrm{g}}$
is met \citep{hon96,man00}, and define the critical accretion rate, $\dot{M}=\dot{M_{\mathrm{cr}}}$,
such that $R_{\mathrm{tr}}=10R_{\mathrm{g}}$ is obtained at $\dot{M}=\dot{M_{\mathrm{cr}}}$.
By further assuming that the accretion disk has a finite outer radius $R_{\mathrm{out}}\equiv1000R_{\mathrm{g}}$, different ranges can be defined by (see also 
Figure 1)
Range I: $R_{\mathrm{tr}}>1000R_{\mathrm{g}}$;
Range II: $1000R_{\mathrm{g}}>R_{\mathrm{tr}}>10R_{\mathrm{g}}$;
and
Range III: $R_{\mathrm{tr}}<10R_{\mathrm{g}}$.

\begin{figure}[h]
\epsscale{.50}
\plotone{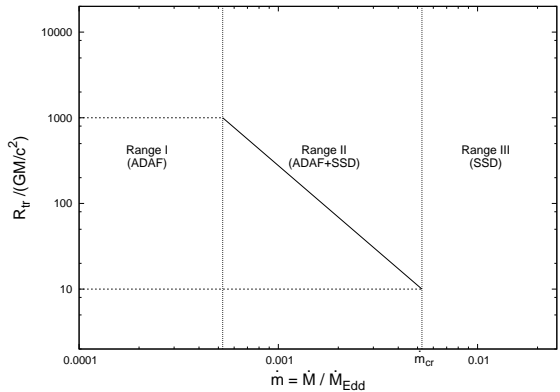}
\caption{ Disk type as a function of the accretion rate $\dot{m}$ in our model. The solid curve
shows the computed value of 
Equation (3) with $\alpha=0.1$.
In our calculation, 
 we assume the accretion disk has a finite outer radius
  $R_{\mathrm{out}}=1000R_{\mathrm{g}}$
  and assume $R_{\mathrm{tr}}=10R_{\mathrm{g}}$ when the accretion rate
$\dot{m}$ equals to the critical accretion rate $\dot{m}_{\mathrm{cr}}$.
Therefore, when $\dot{m}$ has a value within Range I/II/III,
the disk type is of a(n) ADAF/combined disk/Shakura-Sunyaev disk (SSD), respectively.
See the text for more details.\label{fig1}}
\end{figure}

Our strategy to investigate the formation of relativistic jet at different 
disk types is described below.
In contrast to that there is only outflow along the large-scale, \textit{disk-threading} field lines for disk winds
\citep{bla82,sad10},
there are both inflow and outflow along the \textit{hole-threading} field lines.
The inflow (or outflow) is launched when the gravitational force acting
on the plasma is larger (or smaller) than the magnetocentrifugal
forces do.
Such a difference allows us to monitor the outflow properties by the inflow behavior; for an MHD flow along
a hole-threading line, a powerful \textit{outflow} is realized only when the rotational energy is extracted by the \textit{inflow}.
Therefore, the condition for jet formation is to extract the rotational energy
of BH, which can be examined by  solving the 
equation of motion,
namely, that the relativistic Bernoulli equation (BE), of the inflow along the field line (see Section 3 for more details). 

The influences of the disk type lie in how they affect the large-scale field strength ($\bar{B}$)
and the mass loading.
At the injection point of the inflow, $R_{inj}$,
by denoting the superscript {}``ADAF'' and {}``SSD'' as the analytical
solutions of ADAF and SSD, respectively, and assuming that the same
value of $\varepsilon$ can be applied to both ADAF and SSD, 
these two physical properties can be estimated (see Table 1) 
in different region of the computation domain of $\dot{m}$
accordingly.

Range I:\textit{ ADAF type}.--- The large-scale
field at $R_{\mathrm{inj}}$ is estimated by the ADAF surface density,
that is, $\bar{B}_{\mathrm{I}}=B(\varepsilon,R_{\mathrm{inj}},\sigma^{\mathrm{ADAF}})$
by Equation (2). The mass flux per field line at $R_{\mathrm{inj}}$
is estimated by the ADAF solution as well, which gives $\rho^{\mathrm{ADAF}}(R_{\mathrm{inj}})u^{\mathrm{ADAF}}(R_{\mathrm{inj}})/\bar{B_{\mathrm{I}}}$.

Range II:\textit{ Combined disk type}.--- 
The large-scale magnetic field at $R_{\mathrm{inj}}$, which contains an
additional term due to the magnetic field advected from the thin disk to the ADAF,
is estimated by $\bar{B}_{\mathrm{II}}=B(\varepsilon,R_{\mathrm{inj}},\sigma^{A\mathrm{DAF}})+B(\varepsilon,R_{\mathrm{tr}},\sigma^{\mathrm{SSD}})$.
The mass flux per magnetic flux tube at $R_{\mathrm{inj}}$
is computed by $\rho^{\mathrm{ADAF}}(R_{\mathrm{inj}})u^{\mathrm{ADAF}}(R_{\mathrm{inj}})/\bar{B}_{\mathrm{II}}$,
since the injection point is located inside the ADAF.

Range III:\textit{ Thin disk type}.--- The large-scale
field at $R_{\mathrm{inj}}$ is estimated by $\bar{B}_{\mathrm{III}}=B(\varepsilon,R_{\mathrm{inj}},\sigma^{\mathrm{SSD}})$
and the mass flux per field line at $R_{\mathrm{inj}}$ is estimated
by $\rho^{\mathrm{SSD}}(R_{\mathrm{inj}})u^{\mathrm{SSD}}(R_{\mathrm{inj}})/\bar{B}_{\mathrm{III}}$.

\begin{table}
%{>{\centering}m{3cm}cc>{\centering}m{4cm}}
\begin{centering}
%\begin{center}
\caption{Estimating the Field Strength and the Mass Loading at Different Accretion Ranges\label{table1}}
\begin{tabular}{>{\centering}m{2cm}>{\centering}m{3cm}>{\centering}m{5cm}>{\centering}m{5cm}}
%{ccc>{\centering}p{4.5cm}}
\hline
\hline  
& & & \tabularnewline
Range & Disk Type & Field Strength\tablenotemark{a} at the Injection Point & Mass Loading\tablenotemark{b} at the Injection Point\tabularnewline
\hline
& & & \tabularnewline
I & ADAF & $\bar{B}_{\mathrm{I}}=B_{\mathrm{inj}}(\varepsilon,R_{\mathrm{inj}},\sigma^{\mathrm{ADAF}}(R_{\mathrm{inj}}))$ & $\rho^{\mathrm{ADAF}}u^{\mathrm{ADAF}}/\bar{B_{\mathrm{I}}}$\tabularnewline
& & & \tabularnewline
II & ADAF+SSD (Combined) & $\bar{B}_{\mathrm{II}}=B_{\mathrm{inj}}(\varepsilon,R_{\mathrm{inj}},\sigma^{\mathrm{ADAF}}(R_{\mathrm{inj}}))$
$+B_{tr}(\varepsilon,R_{\mathrm{tr}},\sigma^{\mathrm{SSD}}(R_{\mathrm{tr}}))$ & $\rho^{\mathrm{ADAF}}u^{\mathrm{ADAF}}/\bar{B}_{\mathrm{II}}$\tabularnewline
& & & \tabularnewline
III & SSD & $\bar{B}_{\mathrm{III}}=B_{\mathrm{inj}}(\varepsilon,R_{\mathrm{inj}},\sigma^{\mathrm{SSD}}(R_{\mathrm{inj}}))$ & $\rho^{\mathrm{SSD}}u^{\mathrm{SSD}}/\bar{B}_{\mathrm{III}}$\tabularnewline
\hline
\end{tabular}
\par\end{centering}
{$^a$} {See Equation (2).}

{$^b$} {The mass flux per magnetic flux tube.}
\end{table}

In next section,
we summarize necessary concepts of the ideal MHD flow in a BH magnetosphere, including
the relativistic BE and the MHD Penrose process.
The result of our model is provided in Section 4.
%, we can qualitatively examine the extraction
%of the BH rotational energy at various accretion rate.  
%Detailed process to determining
%the location of the Alfven point by solving the BE is provided in the Appendix.

\section{MHD Flows around a Rotating Black Hole}

Stationary and axisymmetric ideal MHD flows in a rotating BH magnetosphere
have been investigated in several studies \citep{cam86a,cam86b,cam87,tak90,hir92}.
In this paper, we adopt the Kerr metric
as the background geometry with signature (-,+,+,+) and use the geometrized
units such that $c=G=M=1$, that is, the length scale is in the unit
of {}`$GM/c^{2}$'. 
Most of the derivations in this section follows the calculations in
\citet{tak90}. Note that a different signature (+,-,-,-) is used in their paper.

\subsection{The Relativistic Bernoulli Equation}
In the Boyer-Lindquist coordinate, the metric
is

\begin{equation}
ds^{2}=-\frac{\Delta-a^{2}\sin^{2}\theta}{\Sigma}dt^{2}-\frac{4ar\sin^{2}\theta}{\Sigma}dtd\phi+\frac{A\sin^{2}\theta}{\Sigma}d\phi^{2}+\frac{\Sigma}{\Delta}dr^{2}+\Sigma d\theta^{2}\;,\end{equation}
where $a\equiv J$, $\Delta\equiv r^{2}-2r+a^{2}$, $\Sigma\equiv r^{2}+a^{2}\cos^{2}\theta$
and $A\equiv(r^{2}+a^{2})^{2}+\bigtriangleup a^{2}\sin^{2}\theta$. 
The ideal MHD condition requires that the electric field vanish in
the fluid's rest frame,

\begin{equation}
\sum_{\mu}F_{\mu\nu}u^{\mu}=F_{\mu\nu}u^{\mu}=0\;,\end{equation}
where $F_{\mu\nu}=A_{\nu,\mu}-A_{\mu,\nu}$ is the electromagnetic
field tensor satisfying the Maxwell's equation, $u^{\nu}$ is the four
velocity of the fluid, $A_{\mu}$ is the electromagnetic vector potential,
and the comma refers to the derivative.

The fluid equation of motion $T_{;\nu}^{\mu\nu}=0$ can be split into
the two components: the {}`relativistic BE' (the poloidal equation) that describes the flow along the field line,
and the {}`Grad-Shafronov equation' (the trans-field equation) that
describes the force balance perpendicular to the field lines, where
the semi-colon refers to the covariant derivative. The BE can be obtained
by projecting the equation of the motion along the field line; however,
it can alternatively derived from the definition of proper time, \begin{equation}
u^{\alpha}u_{\alpha}=-1\;.\end{equation}
The energy-momentum tensor $T^{\mu\nu}=T_{\mathrm{em}}^{\mu\nu}+T_{\mathrm{plasma}}^{\mu\nu}$
consists of two terms. One is the electromagnetic term,

\begin{equation}
T_{\mathrm{em}}^{\mu\nu}=\frac{1}{4\pi}(F^{\mu\alpha}F_{\alpha}^{\nu}-\frac{1}{4}g^{\mu\nu}F^{\alpha\beta}F_{\alpha\beta})\;,\end{equation}
and the other is the plasma term,

\begin{equation}
T_{\mathrm{plasma}}^{\mu\nu}=(P+\rho)u^{\mu}u^{\nu}+g^{\mu\nu}P\;,\end{equation}
where $P$ is the pressure and $\rho$ is the energy density. The proper
number density $n$ obeys the continuity equation \begin{equation}
(nu^{\mu})_{;\mu}=0\;,\end{equation}
and the relativistic specific enthalpy $\mu$ satisfies

\begin{equation}
\mu=m_{\mathrm{p}}+\frac{\gamma}{\gamma-1}\frac{P}{n}=m_{\mathrm{p}}[1+\frac{\gamma}{\gamma-1}\frac{P_{\mathrm{inj}}}{n_{\mathrm{inj}}m_{\mathrm{p}}}(\frac{n}{n_{\mathrm{inj}}})^{\gamma-1}]\end{equation}
where $m_{\mathrm{p}}$ denotes the rest mass energy of the proton,
$\gamma$ is the adiabatic index, and the subscript {}``$\mathrm{inj}$'' is
the quantity evaluated at the particle injection point. The assumption
of an adiabatic flow requires that the entropy along the field line
be constant, which ensures that the term $P_{\mathrm{inj}}/(n_{\mathrm{inj}}m_{\mathrm{p}})$
remains constant along the field line. 

There are four more conserved quantities along the field line: the
angular velocity of the field line, $\Omega_{\mathrm{F}}$ , the particle
flux per unit flux tube, $\eta$, the total energy of the flow per
particle, $E$, and the total angular momentum of the flow per particle,
$L$. They can defined by (cf. Equations(2.3)-(2.6) of Hirotani et al. 1992)

\begin{equation}
\Omega_{\mathrm{F}}=\frac{F_{tr}}{F_{r\phi}}=\frac{F_{t\theta}}{F_{\theta\phi}}\;,\end{equation}

\begin{equation}
\eta=\frac{\sqrt{-g}nu^{r}}{F_{\theta\phi}}=\frac{\sqrt{-g}nu^{\theta}}{F_{\phi r}}=\frac{\sqrt{-g}nu^{t}(\Omega-\Omega_{\mathrm{F}})}{F_{r\theta}}\;,\end{equation}

\begin{equation}
E=-\mu u_{t}-\frac{\Omega_{\mathrm{F}}}{4\pi\eta}\sqrt{-g}F^{r\theta}=-\mu u_{t}+\frac{\Omega_{\mathrm{F}}}{4\pi\eta}B_{\phi}\;,\end{equation}

\begin{equation}
L=\mu u_{\phi}-\frac{1}{4\pi\eta}\sqrt{-g}F^{r\theta}=\mu u_{\phi}+\frac{B_{\phi}}{4\pi\eta}\;,\end{equation}
where $-g\equiv-det\mid g\mid=\Sigma^{2}\sin^{2}\theta$ , $\Omega\equiv u^{\phi}/u^{t}$
denotes the fluid angular velocity, and $B_{\phi}\equiv-\sqrt{-g}F^{r\theta}=-(\Delta/\Sigma)\sin\theta F_{r\theta}$
the toroidal magnetic field seen by a distant static observer with
four velocity $u_{\mathrm{Lab}}^{\nu}=(1,0,0,0)$. One should note
that the sign of $\eta$ is defined by the signs of $u^{r}$ and $F_{\theta\phi}$.
Although the magnetic field is defined by

\begin{equation}
B_{\mu}\equiv\frac{1}{2}\sqrt{-g}\epsilon_{\nu\mu\alpha\beta}F^{\alpha\beta}u_{\mathrm{Lab}}^{\nu}\;;\end{equation}
however, it is convenient to introduce the rescaled poloidal field
\begin{eqnarray}
 & B_{\mathrm{P}}^{2}\equiv B^{A}B_{A}(g_{tt}+g_{t\phi}\Omega_{\mathrm{F}})^{-2}= & \frac{g^{AB}}{g_{t\phi}^{2}-g_{tt}g_{\phi\phi}}(A_{\phi,A}A_{\phi,B}),\end{eqnarray}
where $A$ runs over the poloidal coordinates, $r$ and $\theta$.
Combining Equations (12)-(14), we have

\begin{equation}
F^{r\theta}\frac{\sqrt{-g}}{4\pi\eta}=-\frac{(g_{t\phi}+\Omega_{F}g_{\phi\phi})E+(g_{tt}+\Omega_{F}g_{t\phi})L}{M^{2}-K_{0}}\;,\end{equation}
where

\begin{equation}
K_{0}\equiv-(g_{\phi\phi}\Omega_{\mathrm{F}}^{2}+2g_{t\phi}\Omega_{\mathrm{F}}+g_{tt})\;,\end{equation}
and the poloidal Alfven Mach number is 

\begin{equation}
M^{2}=4\pi\mu n\left(\frac{u_{\mathrm{P}}}{B_{\mathrm{P}}}\right)^{2}=4\pi\mu\eta\left(\frac{u_{\mathrm{P}}}{B_{\mathrm{P}}}\right)=4\pi\mu\eta^{2}/n\;.\end{equation}
The poloidal velocity of the plasma $u_{\mathrm{P}}$ is defined by

\begin{equation}
u_{\mathrm{p}}^{2}\equiv u^{r}u_{r}+u^{\theta}u_{\theta}\;.\end{equation}
Substituting Equation (17) into Equations (13) and (14), we can express the
energy of the fluid, $-\mu u_{t}$, and the angular momentum of the
fluid, $\mu u_{\phi}$, as 

\begin{equation}
-\mu u_{t}=\frac{(g_{tt}+\Omega_{F}g_{t\phi})(E-\Omega_{F}L)+M^{2}E}{M^{2}-K_{0}}\;,\end{equation}

\begin{equation}
\mu u_{\phi}=-\frac{(g_{t\phi}+\Omega_{F}g_{\phi\phi})(E-\Omega_{F}L)-M^{2}L}{M^{2}-K_{0}}\;.\end{equation}
Finally, combining Equations (4), (20), (21), and (22), we obtain the BE
(cf. Equation (17) of Takahashi et al. 1990) ,

\begin{equation}
u_{\mathrm{p}}^{2}+1=(\frac{E}{\mu})^{2}\frac{K_{0}K_{2}-2K_{2}M^{2}-K_{4}M^{4}}{(M^{2}-K_{0})^{2}}\;,\end{equation}
where $K_{2}$ and $K_{4}$ are defined as

\begin{equation}
K_{2}\equiv\left(1-\Omega_{\mathrm{F}}\frac{L}{E}\right)^{2}\;,\end{equation}

\begin{equation}
K_{4}\equiv-\left[g_{\phi\phi}+2g_{t\phi}\frac{L}{E}+g_{tt}\left(\frac{L}{E}\right)^{2}\right]/(g_{t\phi}^{2}-g_{tt}g_{\phi\phi})\;.\end{equation}
The BE, Equation (23), can be rewritten as a polynomial of $u_{P}$ with
the aid of Equations (10) and (19). Note that the order of the polynomial
equation depends on the second term of Equation (10), because particle
number conservation gives $n\varpropto1/u_{\mathrm{p}}$.

\subsection{The Cold Limit}

The cold limit can be adopted
when the second term in Equation (10) is relatively unimportant 
to its first term, i.e., ${P_{\mathrm{inj}}}/{n_{\mathrm{inj}}m_{\mathrm{p}}c^{2}}\rightarrow0$
(in c.g.s unit). 
Equation (10) therefore reduces to $\mu=m_{\mathrm{p}}$ and the BE becomes a
fourth-order polynomial equation of $u_{\mathrm{P}}$. By introducing a
parameter $\zeta\equiv u_{\mathrm{p}}/B_{\mathrm{p}}$, we have

\[
\eta=n\zeta\;,\]

\[
u_{\mathrm{p}}^{2}=\frac{\zeta^{2}}{\Delta\sin^{2}\theta}[g^{rr}(A_{\phi,r})^{2}+g^{\theta\theta}(A_{\phi,\theta})^{2}]\;,\]

\[
B_{\mathrm{p}}^{2}=\frac{1}{\Delta\sin^{2}\theta}[g^{rr}(A_{\phi,r})^{2}+g^{\theta\theta}(A_{\phi,\theta})^{2}]\;,\]

\[
M^{2}=4\pi\mu n\zeta^{2}=4\pi\mu\eta\zeta\;.\]
The cold BE, therefore, can be expressed in terms of $\zeta$\begin{equation}
(\mathcal{A}^{2}\mathcal{B})\zeta^{4}-(2K_{0}\mathcal{A}\mathcal{B})\zeta^{3}+(\mathcal{A}^{2}-\mathcal{A}^{2}\mathcal{C}+K_{0}^{2}\mathcal{B})\zeta^{2}+(2\mathcal{D}\mathcal{A}-2K_{0}\mathcal{A})\zeta^{1}+K_{0}(K_{0}-\mathcal{D})=0\;,\end{equation}
where

\[
\mathcal{A}=4\pi\mu\eta\;,\]

\[
\mathcal{B}=\frac{u_{P}^{2}}{\zeta^{2}}=B_{p}^{2}\;,\]

\[
\mathcal{C}=-(\frac{E}{\mu})^{2}K_{4}\;,\]

\[
\mathcal{D}=(\frac{E}{\mu})^{2}K_{2}\;.\]

Even for a disk whose temperature is comparable with
the virial temperature and the internal energy is comparable to the released 
gravitational energy (e.g., ADAF), the cold limit can still be modestly satisfied. 
This is because the released gravitational binding energy is still a small fraction of the
rest-mass energy.
Therefore, we solve the BE in cold limit, Equation (26), for simplicity and
expect the result is at least qualitatively correct.
In order to specify the coefficients $\mathcal{A}$ and $\mathcal{B}$,
we need to specify the mass loading (mass flux per magnetic flux tube), 
$\mu\eta$, and the strength and geometry of the magnetic
field. It is  provided in the Appendix C how we compute these coefficients.

\subsection{Light Surfaces and the Separation Point}
\begin{figure}[h]
\epsscale{.50}
\plotone{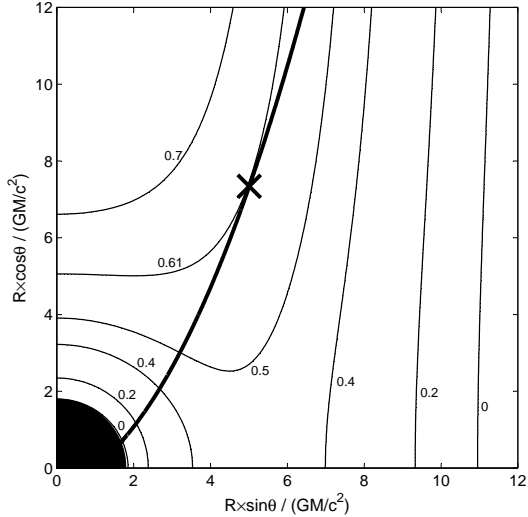}
%\plottwo{f2.eps}{f2_color.eps}
\caption{Illustration of the separation point of a field line and
contours of $K_{0}$ around a black hole with spin parameter $a=0.6$
and $\Omega_{F}=0.5\Omega_{H}$. Light surfaces are defined by $K_{0}=0$.
The Alfven point of a physical MHD flow must
be located inside the two (inner and outer) light surfaces, because $M^{2}|_{r_{A}}=K_{0}\mid_{r_{A}}\geq0$.
Black region represents the BH. For a field line threading the
BH (black solid line), the separation point is determined by $K_{0}^{'}=0$
(indicated by a cross). 
See the text for more details.\label{fig2}}
\end{figure}
There are two light surfaces, which are defined by $K_{0}=0$, in
a BH magnetosphere. The outer light surface is formed by the centrifugal
force in the same manner as the light cylinder in a pulsar magnetosphere,
and the inner light surface is, on the other hand, formed by the strong
gravity of the BH and the rotation of the magnetic field. 

If a plasma starts with a negligible poloidal velocity, the inflow and
outflow along the field line separate at the point $r_{\mathrm{s}}$
where the gravitational force balances with the magnetocentrifugal
forces. This point is called the {}``separation point'' (or the {}``stagnation
point'') and defined by $K_{0}^{'}=0$, which leads to $\left(\ln u_{P}\right)^{'}=0$,
where the prime denotes the derivative along the flow line. That is, if a plasma is injected at $r=r_{\mathrm{inj}}$
with $r_{\mathrm{inj}}<r_{\mathrm{s}}$ (or with $r_{\mathrm{inj}}>r_{\mathrm{s}}$),
it will be accelerated inward (or outward), subsequently passing through
the Alfven point $r=r_{\mathrm{A}}$ and the inner (or outer)
light surface. Figure 3 shows the contours of $K_{0}$ around a BH
 with spin parameter $a=0.6$. Note that, at the injection point,
the BE reduces to

\begin{equation}
E-\Omega_{\mathrm{F}}L=\mu\sqrt{K_{0}}\mid_{r_{\mathrm{inj}}}\;.\end{equation}

\subsection{Critical Points}

The critical points for an MHD flow can be found by differentiating
Equation (23) along the flow line, $\left(\ln u_{P}\right)^{'}=N/D$,
where explicit expressions of $N$ and $D$ are given in \citet{cam86b} and
\citet{tak90}. For a general MHD inflow, there are three critical points at which
both $D$ and $N$ vanish: the {}`slow-magnetosonic point', the
{}`Alfven point', and the {}`fast-magnetosonic point'; the poloidal
velocity $u_{\mathrm{p}}$ there matches the slow-magnetosonic velocity,
the Alfven velocity, and the fast-magnetosonic velocity, respectively.
In the cold limit, the slow-magnetosonic velocity reduces to zero.
Due to the causality requirement at the event horizon, any MHD inflow
must pass through the fast-magnetosonic point, after passing through
the Alfven point. 

At the Alfven point $r=r_{\mathrm{A}}$, the Mach number equals $K_{0}$, 

\begin{equation}
M^{2}|_{r_{\mathrm{A}}}=K_{0}|_{r_{\mathrm{A}}}\;.\end{equation}
Note that the requirement of $M^{2}|_{r_{\mathrm{A}}}=K_{0}\mid_{r_{\mathrm{A}}}\geq0$
implies that $r_{\mathrm{A}}$ must be located between these two light
surfaces. Besides, it is convenient to express $E$ and $L$ in terms
of the quantities evaluated at the Alfven point. Since both $D$ and
$N$ automatically vanish at the Alfven point, no additional constraints
are imposed on the MHD flow at the Alfven point. Therefore, by requiring
that the numerators in Equations (21)-(23) vanish at $r=r_{\mathrm{A}}$,
we obtain

\begin{equation}
E=-\frac{(g_{tt}+\Omega_{\mathrm{F}}g_{t\phi})(E-\Omega_{F}L)}{M^{2}}\mid_{r_{\mathrm{A}}\;},\end{equation}

\begin{equation}
L=\frac{(g_{t\phi}+\Omega_{\mathrm{F}}g_{\phi\phi})(E-\Omega_{F}L)}{M^{2}}\mid_{r_{\mathrm{A}}}\;,\end{equation}
and

\begin{equation}
K_{2}\mid_{r_{\mathrm{A}}}=-K_{2}K_{0}\mid_{r_{\mathrm{A}}}\;.\end{equation}
Combining Equations (27), (29), and (30), we can express the conserved quantities
$E$ and $L$ in terms of $r_{\mathrm{inj}}$ and $r_{\mathrm{A}}$. 

After the MHD inflow pass through the Alfven point, $D$ vanishes
again in $\left(\ln u_{P}\right)^{'}=N/D$ at the fast-magnetosonic
point; thus, $N$ should also vanish there. We should notice here
that both $D=0$ and $N=0$ can be satisfied at the fast-magnetosonic
point only for a specific combination of the conserved quantities,
$\Omega_{\mathrm{F}}$, $E$ , $L$, and $\eta$. In previous works, e.g.,
\citet{cam86a} and \citet{tak90}, the authors gave $\Omega_{\mathrm{F}}$, $E$,
and $L$ (or equivalently, $\Omega_{\mathrm{F}}$, $r_{\mathrm{inj}}$,
and $r_{\mathrm{A}}$) and search $\eta$ to select the trajectory
that pass through the fast-magnetosonic point. In this work, we instead
give $\Omega_{\mathrm{F}}$, $r_{\mathrm{inj}}$, and $\eta$, considering
plasma injection from the innermost region of the accretion disk,
and solve for $r_{\mathrm{A}}$ (or equivalently, $E$).
As explained next, whether the rotational energy of the BH 
is extracted outward can be determined by the  location of $r_{\mathrm{A}}$.

\subsection{Negative Energy Flow}

\begin{figure}[h]
\epsscale{.60}
\plotone{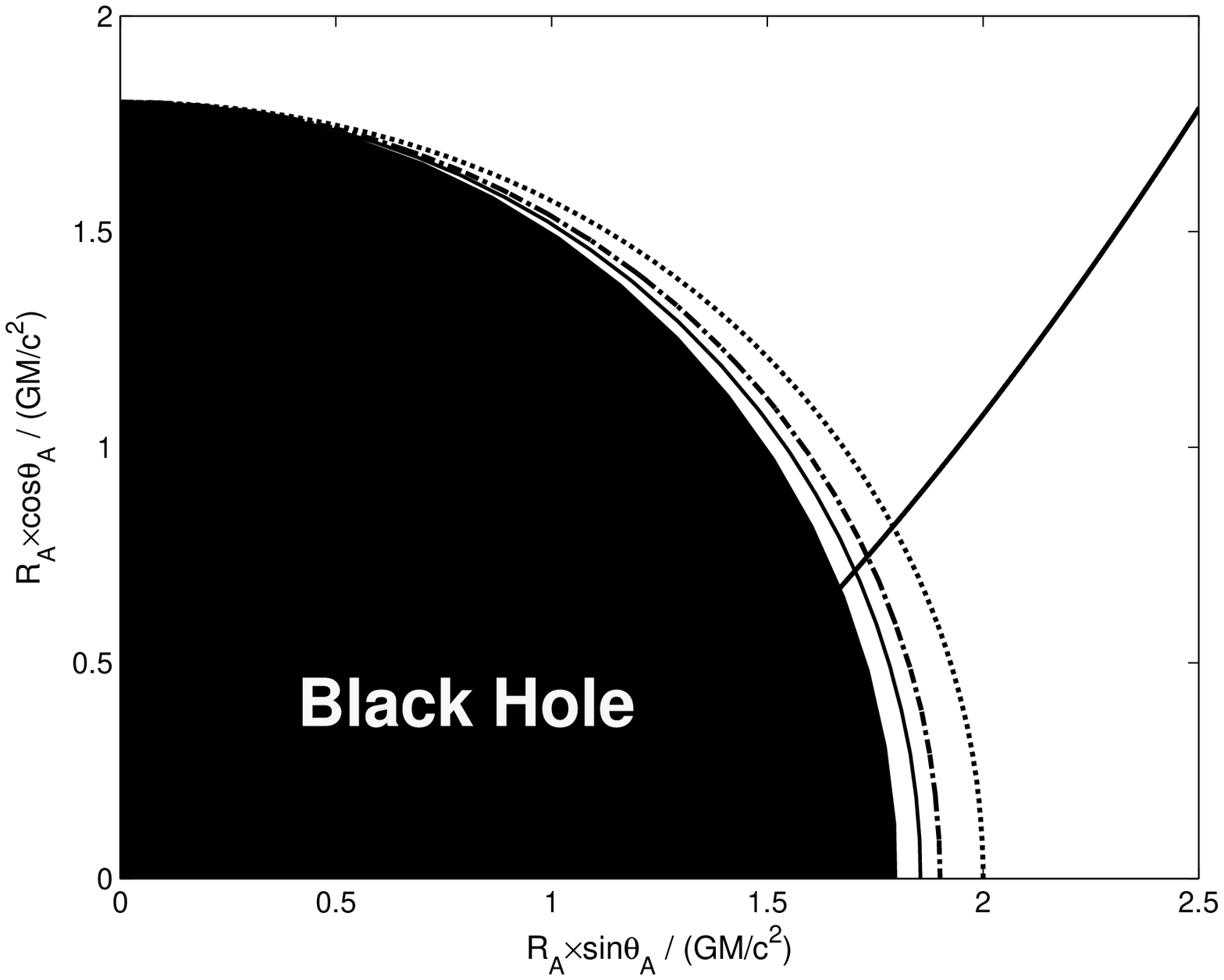}
\caption{Spatial position of the Alfven point that gives a negative
energy flow for the case of $\Omega_{F}=0.5\Omega_{H}$ and $a=0.6$.
The dash-dot line, which is defined by $g_{tt}+\Omega_{F}g_{t\phi}=0$,
 is located between the inner light surface (thin solid
line) and the static limit (dotted line), the latter which defines
the outer boundary of the ergosphere. An MHD flow has a negative energy
if the Alfven point is located within the negative-energy region,
which is defined as the region between the dash-dot and the thin solid
lines. The magnetic field threading the hole surface in Figure 2 is
also shown (thick sold line). See also Figure 6 of \citet{tak90}.\label{fig3}}
\end{figure}

The sign of $E$ can be solely determined by $r_{\mathrm{A}}$ via
Equation (29) because it follows form Equation (27) that $E-\Omega_{\mathrm{F}}L>0$
holds for any MHD flow starting from a position between the two light
surfaces with a vanishing poloidal velocity. Note that $E$ becomes
negative if $(g_{tt}+\Omega_{\mathrm{F}}g_{t\phi})|_{r_{A}}>0$ . Thus, near
the BH, a spatial region called the {}``negative energy region'',
which satisfies $(g_{tt}+\Omega_{\mathrm{F}}g_{t\phi})>0$, can be
defined \citep{tak90}. If the Alfven point resides in this region, the MHD inflow
has $E<0$ (which also implies $L<0$ because of $E-\Omega_{\mathrm{F}}L>0$;
whereas $L<0$ does not imply $E>0$). 

A spinning BH differs from a static BH by an additional structure called {}``ergosphere'' 
(which is defined by $g_{tt}>0$), a region inside which any particle must orbit in the same direction 
of the hole spin. Being locating outside the horizon, the ergosphere can be viewed as 
the region that stores the rotation energy of a BH.
Note that the negative energy
region exists only when $\Omega_{H}>\Omega_{F}>0$, 
 and it is always located inside
the ergosphere (see also Figure 3). It is also
noteworthy that a negative energy MHD inflow can be launched from
a region outside the ergosphere.

\subsection{Outward Energy Flux and the MHD Penrose Process}

\begin{table}
\label{table2}
%{>{\centering}m{3cm}cc>{\centering}m{4cm}}
\begin{centering}
%\begin{center}
\caption{Inflow Energy as the Criterion for Ergospheric Jet Launch}
\begin{tabular}{>{\centering}m{6cm}>{\centering}m{3cm}>{\centering}m{3cm}>{\centering}m{3cm}}
\hline 
\hline
 & & & \tabularnewline
\multicolumn{1}{>{\centering}p{5cm}}{Inflow Property} & Infow Energy\tablenotemark{b}& Inflow Energy Flux\tablenotemark{c}  & Jet Launch?\tablenotemark{d} \tabularnewline
\hline
 & & & \tabularnewline
plasma dominated ($\mathcal{E}_{\mathrm{em}}^{r}<|\mathcal{E}_{\mathrm{plasma}}^{r}|$)\tablenotemark{a}  & $E>0$ & $\mathcal{E}^{r}<0$ & NO\tabularnewline
 & & & \tabularnewline
magnetically dominated ($\mathcal{E}_{\mathrm{em}}^{r}>|\mathcal{E}_{\mathrm{plasma}}^{r}|$)  & $E<0$ & $\mathcal{E}^{r}>0$ & YES\tabularnewline
\hline
\end{tabular}
\par\end{centering}
%\tablecomments{\\a}
{$^a$} {$\mathcal{E}_{\mathrm{em}}^{r}>0$ and $\mathcal{E}_{\mathrm{plasma}}^{r}<0$ are usually satisfied. See the text for more details.}

{$^b$} {The total energy $E$, Equation (13), contains both the electromagnetic and the plasma contribution.}

{$^c$} {$\mathcal{E}^{r}=n\, E\, u^{r}=\mathcal{E}_{\mathrm{em}}^{r}+\mathcal{E}_{\mathrm{plasma}}^{r}$, Equation (32), where $u^{r}<0$ holds for an inflow.}

{$^d$} {A relativistic jet, which has $u^{r}>0$, $E>0$, and hence $\mathcal{E}^{r}>0$, can be launched only when it connects to an inflow with $\mathcal{E}^{r}>0$}

\end{table}

The outward energy flux of the flow $\mathcal{E}^{r}$ can be split
into the electromagnetic part $\mathcal{E}_{\mathrm{em}}^{r}$ and
the fluid part $\mathcal{E}_{\mathrm{plasma}}^{r}$, 

\begin{equation}
\mathcal{E}^{r}\equiv-T_{t}^{r}=nEu^{r}=\mathcal{E}_{\mathrm{em}}^{r}+\mathcal{E}_{\mathrm{plasma}}^{r}\;,\end{equation}
where,

\begin{equation}
\mathcal{E}_{\mathrm{em}}^{r}\equiv-\frac{1}{4\pi}F^{r\theta}F_{t\theta}=-\frac{\Omega_{F}}{4\pi}F^{r\theta}A_{\phi,\theta}=\frac{\Omega_{F}}{4\pi}\frac{B_{\phi}}{\Sigma\sin\theta}A_{\phi,\theta}.\end{equation}

\begin{equation}
\mathcal{E}_{\mathrm{plasma}}^{r}\equiv-n\mu u_{t}u^{r}\;.\end{equation}
Therefore, a negative energy inflow ($E<0$ and $u^{r}<0$) results in a positive
outward energy flux ($\mathcal{E}^{r}>0$), that is, the BH rotational
energy is extracted.  Note that $\mathcal{E}_{\mathrm{em}}^{r}>0$ usually holds because
its sufficient condition, $0<\Omega_{F}<\Omega_{H}$, is satisfied
under normal conditions \citep{mac82}, where $\Omega_{H}$  refer
to the rotational angular velocities of the BH. 
Note also that the plasmas usually cannot extract BH's
energy (i.e., $\mathcal{E}_{\mathrm{plasma}}^{r}<0$ usually holds) 
because of their rest-mass contribution, unless they are in a negative
energy orbit, which is rare.

In a magnetically dominated
magnetosphere ( $\mathcal{E}_{\mathrm{em}}^{r}> \mathcal{E}_{\mathrm{plasma}}^{r}$), because $\mathcal{E}^{r}$
is continuous across the separation point,  a stationary outflow solution ($u^{r}>0$,
$E>0$ and $\mathcal{E}^{r}>0$) is possible only when it connects
a negative-energy inflow ( $E<0$ and $u^{r}<0$, such that $\mathcal{E}^{r}>0$).
Consequently, a stationary MHD jet from a BH ergosphere
( a   {}``ergospheric jet'') 
can be launched in a magnetically dominated magnetosphere
 only when the MHD Penrose process, $\mathcal{E}^{r}>0$ , operates. 
If a magnetic dominance breaks down, $\mathcal{E}^{r}$ can be discontinuous
across the separation point, that is, an outflow ($\mathcal{E}^{r}>0$)
can connect to a positive-energy inflow ($\mathcal{E}^{r}<0$), owing
to the energy supplied by the accreting plasma near the separation
point. However, this kind of outflow, similar to the disk wind, will
not be accelerated into relativistic energies and therefore is not
of interest in the present paper. To summarize, for an MHD flow
along the hole-threading field line, 
a stationary outflow with positive outward energy flux
($\mathcal{E}^{r}>0$) should connect an inflow solution with $\mathcal{E}^{r}>0$
at the separation point, provided that the injected energy flux by
accretion is small compared to the Poynting flux. Thus, we can investigate
the activity of a Poynting-flux-dominated jet (or an outflow) by examining
the inflow energy, $E$ (see also Table 2).

It should be noted here that the term {}``MHD Penrose
process'' refers to the case of $\mathcal{E}^{r}>0$ (remember $E$ is the {}`total' 
energy, consisting both the electromagnetic plus plasma energy of the flow), as initially
proposed by \citet{tak90}; however, it was used to indicate $\mathcal{E}_{\mathrm{plasma}}^{r}>0$
(i.e., a negative energy orbit of the particle) in some previous studies
\citep{hir92,koi02,sem04,kom05}. The latter case is possible, but can only
be transient \citep{koi02,kom05}. In comparison, in the original idea of the {}`Penrose process' \citep{pen69}, only
the particle contribution ($\mathcal{E}^{r}=\mathcal{E}_{\mathrm{plasma}}^{r}>0$)
was considered; whereas in the Blandford-Znajek process \citep{bla77},
only the electromagnetic contribution ($\mathcal{E}^{r}=\mathcal{E}_{\mathrm{em}}^{r}>0$)
was considered.

\section{Result}

\begin{figure}
\epsscale{.9}
\plotone{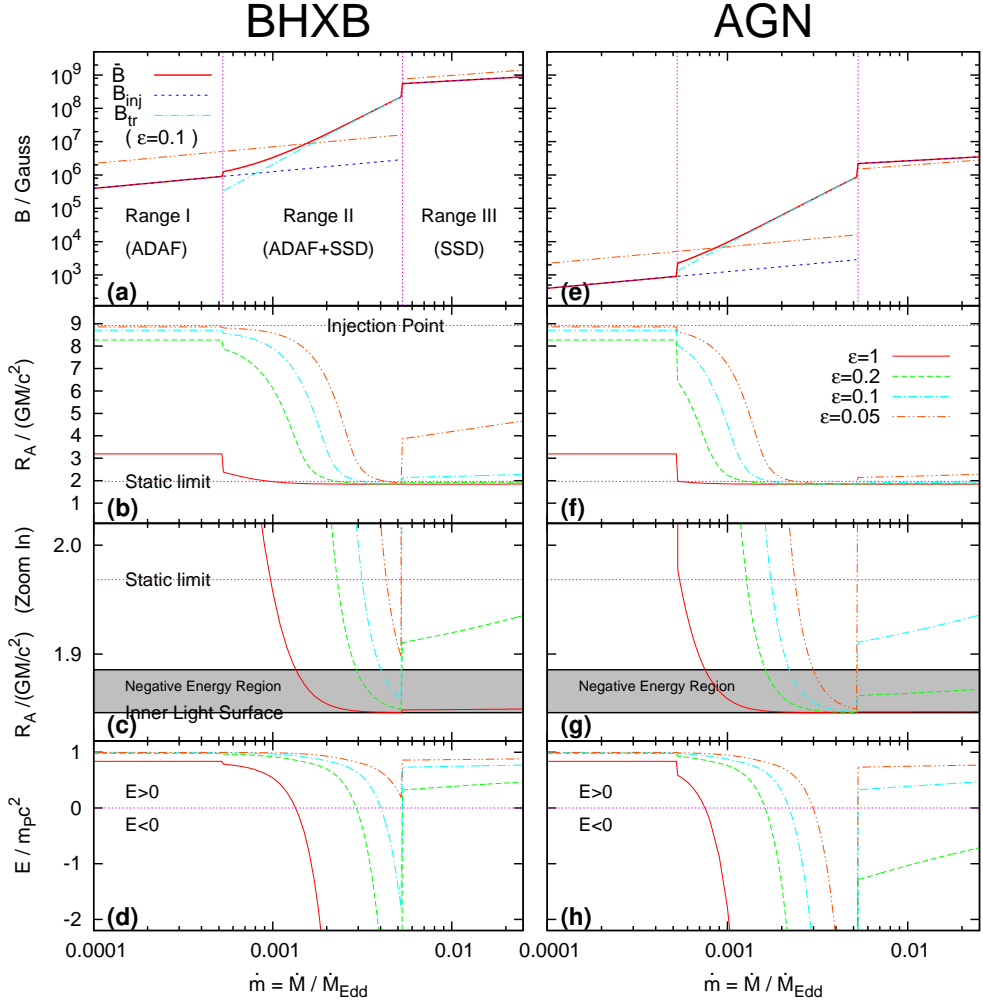}
%\plotone{fig4_uncolored.eps}
\caption{Extraction of BH rotational energy by MHD inflow
(left column: for a BH mass of $10M_{\odot}$; right column: $10^{6}M_{\odot}$).
The domain of accretion rate $\dot{m}$ is divided into three Ranges
(I, II and III) according to the disk types. The parameter $\varepsilon$
($\leq1$) is the squared root of the ratio between the large-scale
magnetic field energy and the disk gravitational binding energy; that
is, $\varepsilon$ denotes the ability for a disk to confine the large-scale
field. (a) and (e) Large-scale magnetic
field strength $\bar{B}$ (solid line) at the injection point. In
Ranges I and III, only $B_{\mathrm{inj}}$ (dashed line) contributes
to $\bar{B}$, where $B_{\mathrm{inj}}$ denotes the field strength
evaluated at the injection point. In Range II, an additional
term $B_{\mathrm{tr}}$ (dash-dotted line) is added so that the fields
advecting from the outer SSD into the inner
ADAF region may be taken into
account, where $B_{\mathrm{tr}}$ denotes the field strength evaluated
at the transition radius of the SSD and the ADAF (see Section 2 for more details). For comparison, we also
plot the strength of local magnetic field when the plasma beta becomes
1 at the injection point as the dash-dot-dot line. To avoid complexity,
only the case of $\varepsilon=0.1$ is shown here, while $\varepsilon=1$,
$0.2$, $0.1$ and $0.05$ are shown in the rest of panels. (b) and (f) The location of the Alfven point
$R_{\mathrm{A}}$. (c) and (g) Zoom-in
plot of the second panel, (b) and (f), near the ergosphere. The static limit indicates
the outer boundary of the ergosphere. The flow energy becomes negative
if $R_{\mathrm{A}}$ is located inside the {}`negative energy region'
indicated by the shaded region (see also Figure 3).  (d) and (h)
Total energy (consisting of plasma and magnetic energies) per particle
in unit of particle rest mass energy, $m_{\mathrm{p}}c^{2}$. Unless
$E\approx m_{p}c^{2}$, the magnetosphere becomes magnetically dominated.
\label{fig4}}

\end{figure}

We choose representative parameters to  demonstrate the jet behavior 
at different disk type qualitatively.
Because the extraction of the rotational energy of BH is mainly related to whether the BH magnetosphere
is magnetically dominated or not, the change of the parameters will
result in only minor modification of the result in our model.

We adopt the dimensionless BH spin parameter
to be $0.6$ and the angular velocity of the field, $\Omega_{\mathrm{F}}$,
to be half of the angular velocity of the hole, $\Omega_{\mathrm{H}}$ (i.e.,
$\Omega_{\mathrm{F}}=0.5\Omega_{\mathrm{H}}=0.0833$). 
The MHD inflow along the field line that intersects the
event horizon at a modest latitude ($\theta_{\mathrm{H}}\sim1.1868\mathrm{rad}$,
i.e., about $68^{\circ}$ from the pole;
see Figures 2 and 3 for this field line) is considered,
and the inflow is assumed to be launched at the separation point
(that is, the injection point is located at
the separation point). 
In addition, to investigate the importance of the BH mass, 
we consider separately the
BHXB and AGN cases.
For BHXB, we assume $M_{\mathrm{BH}}=10M_{\odot}$, whereas for AGN, $M_{\mathrm{BH}}=10^{6}M_{\odot}$. 
Furthermore, to explore the dependence of 
the ability of the disk in storing the
large-scale magnetic field, we
adopt different values 
of $\varepsilon$ ($\varepsilon=1,0.2,0.1$,
and $0.05$).

After starting from the injection point, the
accreting MHD plasma is accelerated inward, passing thorough the Alfven
point, $R_{\mathrm{A}}$, and the fast-magnetosonic point,
and finally enter the BH. The inflow energy $E$
can be computed by imposing the fast-magnetosonic condition
once the injection point, the angular velocity of the field and the
 mass loading are specified.
However, to discuss the conditions for jet launching ($\mathcal{E}^{r}>0$)
and quenching ($\mathcal{E}^{r}<0$), it is adequate to examine
the \textit{sign} of $E$, 
which is uniquely specified by the location of $R_{\mathrm{A}}$.
Specifically, if $R_{\mathrm{A}}$ resides in the  negative
energy region, the MHD inflow has a negative energy
($E<0$) and the MHD Penrose process operates. Therefore, by solving
$R_{\mathrm{A}}$ from the BE
for an MHD inflow onto a BH, we can investigate whether the energy
extraction takes place as a function of $\dot{m}$. 
Detailed computation steps are provided in the Appendix.

Let us first consider the BHXB case with the BH mass, $M_{\mathrm{BH}}=10M_{\odot}$
(Figures 4(a)-(d)).
Figure 4(a) shows how the large-scale field strength at the injection point varies
with the accretion rate.
It is important to note that in a combined disk (Region II), the surface
density of the disk does
not monotonically increase inward but has a local maximum at the
transition radius, $R_{tr}$. Since an SSD can confine more large-scale
field than an ADAF because of its greater surface density for a fixed $\varepsilon$,
the large-scale field arrested in the inner ADAF for a combined disk (solid line) is
relatively stronger than what the ADAF  alone would bring
into the BH (dashed line). 
The locations of $R_{\mathrm{A}}$  for
four discrete values of $\varepsilon$ are shown in Figures 4(b) and 4(c); the solid, dashed, dash-dotted,
and dash-dot-dot lines correspond to $\varepsilon=1$, $0.2$, $0.1$,
and $0.05$, respectively. It follows from the zoom-in figure (Figure
4(c)) that $R_{\mathrm{A}}$ enters the negative energy region (indicated
by the shaded region) in Range II for $\varepsilon=1$, $0.2$, and
$0.1$, indicating that the MHD Penrose process turns on in a combined
disk. 
This is because, near the horizon, the relatively strong magnetic field
(which is mainly provided by the outer SSD) and the relatively small mass loading
(which is provided by the inner ADAF) result in a magnetically dominated
environment. In comparison,
a magnetically dominated environment near the horizon is relatively harder to achieved 
when the disk is an ADAF (Range I) or an SSD (Range III), 
because it is the same type of disk
responsible for both the confinement of the large-scale field and the loading
of the plasma. Such disk-jet coupling provides the reason why observed relativistic jet is launched when
  $\dot{m}$ is near (but smaller)
the value $0.01$,
but quenched when $\dot{m}$ exceeds $\sim0.01$.
In addition, since $\bar{B}$ increases with increasing $\dot{m}$,
$E$ increases negatively (i.e., the rotational energy of BH is more efficiently
extracted outwards) with increasing $\dot{m}$ once $R_{A}$ enters
the negative energy region in Range II (Figure 4(d)). This is because
the BH magnetosphere becomes more magnetically dominated when
the transition radius decreases with $\dot{m}$, resulting from that the difference between
the surface density of the outer SSD and inner ADAF becomes more significant.

Next, let us consider the AGN case with $M_{\mathrm{BH}}=10^{6}M_{\odot}$ (Figures
4(e)-(h)). Although the solutions look qualitatively similar, there are
three major differences from the BHXB case. First, comparing Figures 4(b), (c), (f) and
(g), we find that for an AGN, $E<0$ is realized in a wider
range of $\dot{m}$ for a fixed $\varepsilon$ or in a wider range
of $\varepsilon$ for a fixed $\dot{m}$. Second, comparing
Figures 4(d) and (h), we find that an AGN can extract more BH rotational
energy per plasma particle than a BHXB can. Third, in an AGN, the MHD
Penrose process works also in an SSD (i.e., Range III) for a relatively
greater $\varepsilon$ (e.g., $\varepsilon>0.2$). However, the jet
power decreases with increasing $\dot{m}$ in an SSD (i.e., Range III),
in contrast to the case of a combined disk (i.e., Range II).
The above comparison can help to understand the observed properties of accreting BH
in different masses (see below).

\section{Discussion}
\subsection{Universal Paradigm of Black Hole Disk-Jet Coupling at Low Accretion Rate}
\begin{figure}[h]
\epsscale{.80}
\plotone{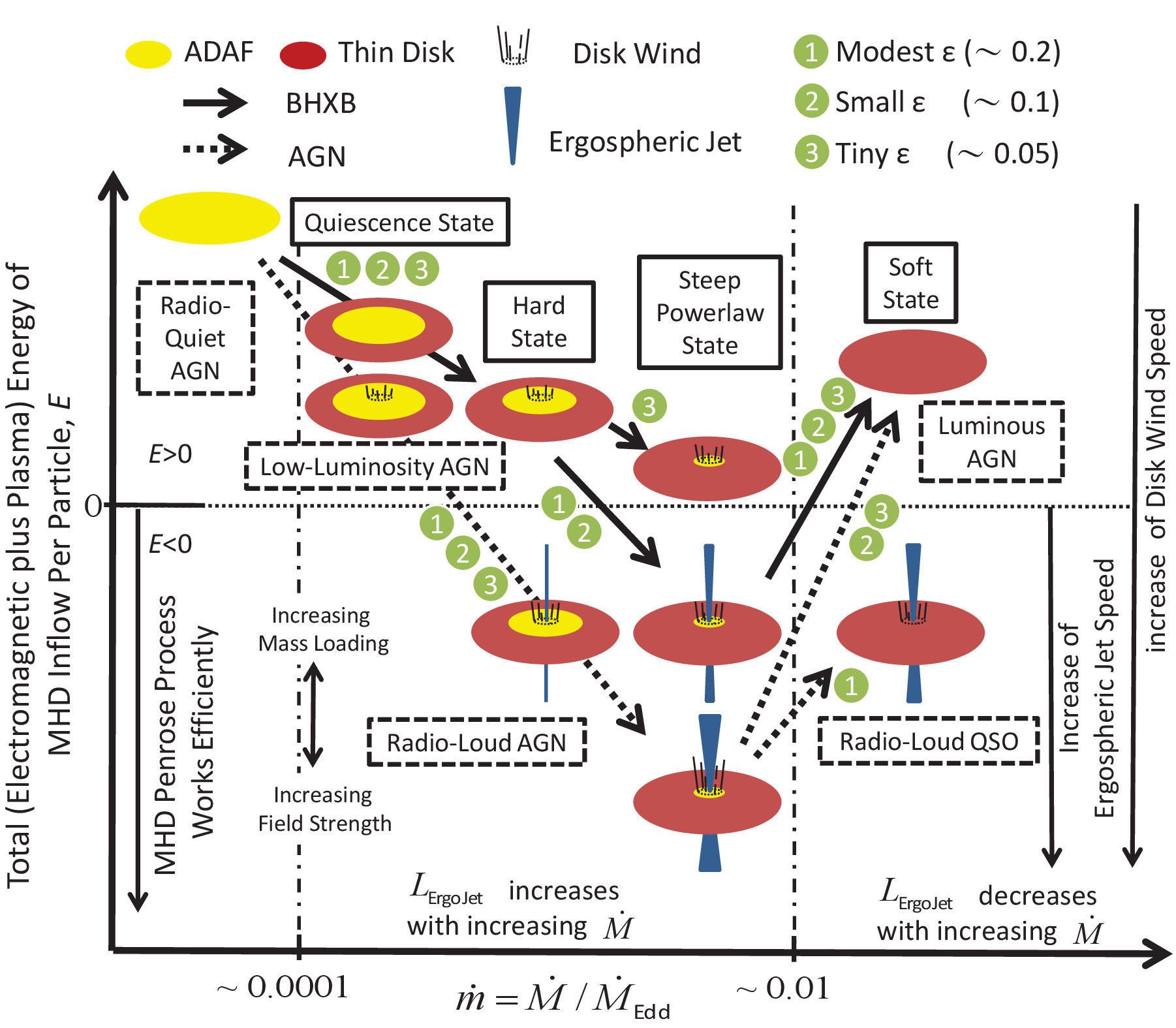}
\label{fig5}
\caption{Universal paradigm of the disk-jet coupling for both
BHXBs and AGNs. As $\dot{m}$ increases, the disk-jet system transits
along the solid arrows for BHXBs and along the dashed arrows for AGNs.
The tracks of transit are depicted for three representative values
of $\varepsilon$, according to the result in Figure 4. Puzzling observational
disk-jet features, such as why radio emission is quenched beyond a
certain $\dot{m}$ and why the radio luminosity decreases (or increases)
with increasing X-ray luminosity when the system is luminous (or less
luminous), are explained. See the text for more details.\\ \\}
\end{figure}

The results in Section 4 can offer the key to an understanding of the observed
connections between the disks and the jets for BHXBs and AGNs. A universal
paradigm of disk-jet couplings on the $\dot{m}-E$ plane (c.f. Figures 4d \& 4h)
is diagrammatically
presented in Figure 5. Transitions of disk-jet states are indicated
by the arrows for three representative values of $\varepsilon$. Since
the terminal speed of an MHD outflow is roughly determined by the 
magnetic dominance at the launching point, the more magnetically dominated the
disk innermost region becomes (i.e., the more a system moves downward
on the $\dot{m}-E$ plane), the greater speed an MHD outflow
 achieves. This discussion is valid for both
the disk wind near the BH and the ergospheric jet.
The two downward arrows in the right edge of this figure indicate
how the terminal speed changes for a disk wind and a relativistic jet. Note
that the ergospheric jet can only be launched when $E<0$ (i.e., in
the lower half of the diagram).

The transitions of disk-jet states for BHXBs, which are indicated
by the solid arrows, explain the observed features of the BHXB disk-jet
couplings.
\citet{fen04} show that two type of outflows associated 
with the X-ray spectral types 
are required to explain 
the systematic properties of the radio emission of BHXBs. The first type
is a steady outflow observed when the BHXB is in a hard state (HS),
while the second type is a transient, ballistic, powerful and relativistic jet when
the source is in a steep power-law (SPL)
state  and being switching
from the HS to the soft state (SS). 
We interpret that the steady outflow (first type) is associated with the disk wind
and the transient relativistic jet (second type) is associated with the ergospheric jet
 ejected for a modest or small $\varepsilon$. 
Note that, as shown in the paradigm,
when the entire disk become an SSD, 
not only that
the ergospheric jet is turned off 
but also that the disk wind 
weakens,  which explains the sudden
drop of observed radio luminosity when the state transits to the SS.
In addition,
 the energy of the jet per particle, $|E|$, as well
as the ergospheric jet power, $L_{\mathrm{ErgoJet}}\equiv\int\mathcal{E}^{r}dS$,
increase with increasing $\dot{m}$ in a combined disk, where the
surface integral $\int dS$ is carried out on a closed surface (e.g.,
the horizon); this likely corresponds to the observational fact that
both the Lorentz factor of a jet and its radio luminosity rapidly increase
with increasing $\dot{m}$ before the jet is quenched.
When the BH is surrounded by a combined disk and the MHD Penrose process takes place, 
the relatively strong field for the inner ADAF
can result in a configuration which has
been studied in previous simulations \citep{nar03,igu03,igu08,pun09,mck12},
here we use the name \textit{magnetcially choked accretion flow (MCAF)}\footnote{An MCAF is formed when  the accumulated 
magnetic field near the BH reaches the equipartition level. 
The strong magnetic field is
able to support (or disrupt) the disk at certain radius $R_{\rm{m}}$, where  $\varepsilon=1$. 
Although slowed down by the magnetospheric barrier near $R_{\rm{m}}$, 
accreting plasma can still accrete inward in the form of
highly non axisymmetry, irregular streams due to magnetic Rayleigh-Taylor instability \citep{igu03,pun09,mck12}.
We stress that the magnetically dominated magnetosphere developed when 
the BH surrounded by a combined disk can result in 
 an MCAF or other conceptually similar accretion flow.} according to 
\citet{mck12}.
 Accompanying by the jet formation, the formation of an MCAF
can also explain some observation features when the source is in SPL.
For an MCAF, because the slowed down flow can 
earn more time to release their heat energy via radiation,
the efficiency for the flow to convert mass to 
radiation energy increases \citep{nar03}.
This can be the reason why the disk luminosity in SPL state is higher than
that in HS or SS \citep{fen04}.
In addition, 
the high-frequency quasi-periodic oscillation (QPO) feature 
observed at  SPL  state
may be related to the jet-disk QPO mechanisms found by
\citet{mck12}.

For AGNs, the transitions are indicated by the dashed arrows, which
imply the following disk-jet couplings. First, an efficient MHD Penrose
process (i.e., a greater $|E|$) is viable at a relatively smaller
$\dot{m}$ in an AGN compared to a BHXB. Since the SSD exists only
in the outer region when $\dot{m}$ is small, it is consistent with
the observational fact that radio-loud AGNs favor a spectrum without
a UV bump \citep{kaw09}, which possibly arises in the inner edge of
an SSD. Second, the MHD Penrose process works more efficiently with
increasing BH mass. This conclusion explains why AGN jets show higher
Lorentz factors than BHXB jets. Moreover, it is also consistent with
the observational fact \citep{liu06,chi11} that most radio-loud AGN host
BHs with larger masses ($\sim10^{8}M_{\odot}$), while radio-quiet
AGNs host BHs with no obvious mass constraints. Third, although a
BHXB jet disappears when the disk transits from a combined disk to
a SSD, an AGN jet could still be launched from the ergosphere, because
the SSD in an AGN tends to be more magnetically dominated than in a BHXB
owing to a smaller mass loading, $\rho u^{r}/\bar{B}$, around a more massive
BH. We interpret that these ergospheric jets ejected from the inner
edge of the thin disks in AGNs result in (at least part of) the radio-loud quasi-stellar objects;
this forms
a striking contrast to BHXBs from which no strong or steady radio
emission has been detected in their SS. Estimated from the
dynamical timescale, the characteristic 
time scale for an accreting BH system to sustain the relativistic jet, $t$,
is scaled by the central BH mass with the relation $t\propto M_{\rm{BH}}$.
Therefore, with $t$ equals to days for the relativistic jet from a BHXB with $M_{\rm{BH}}\sim10M_{\odot}$, relativistic jets from an AGN with 
$M_{\rm{BH}}\sim10^{8}M_{\odot}$ may last for $\sim10^{5}$ years.

Note also that the relation between the accretion rate,
$\dot{m}$, and the ergospheric jet power, $L_{\mathrm{ErgoJet}}$,
inverses in Ranges II and III. That is, $L_{\mathrm{ErgoJet}}$ increases
(or decreases) with increasing $\dot{m}$, when $\dot{m}$ is relatively
low (or high) and the disk is a combined disk (or an SSD). Provided
that the X-ray luminosity indicates the disk accretion rate, these
conclusions naturally explain the empirical relation for both BHXBs
and AGNs: the radio luminosity increases with increasing X-ray luminosity
when the bolometric luminosity, $L$ , is small, i.e., when 
$L/L_{\mathrm{Edd}}=\dot{M}/\dot{M}_{\mathrm{Edd}}\sim1\% $ \citep{mer03,fal04} , while the radio
luminosity decreases with increasing X-ray luminosity when some BH
systems are luminous, i.e., when
$L/L_{\mathrm{Edd}}\sim10\% $\citep{kin11} .

\subsection{The Estimated Speed of Ergospheric Jet}
At large distances, the spacetime is described by the Minkowski metric; thus, 
the Lorentz factor of the jet, $\Gamma$,  is defined by $\Gamma\equiv u^{t}$. Because 
$E$ is conserved quantitatively along the field lines and because the second
term of Equation (13) becomes negligibly small  compared to the first term at large distances, 
we obtain 
\begin{equation}
 \Gamma=\frac{E_{\rm{out}}}{\mu}\;, 
\end{equation}
where $E_{\rm{out}}$ denotes the total energy of the MHD \textit{outflow}.

We can further estimate the terminal jet speed, by linking $E_{\rm{out}}$ with 
the total energy of the MHD \textit{inflow},
$E_{\rm{in}}$, as:
\begin{equation}
E_{\rm{out}}
=\frac{n_{\rm{in}}u^{r}_{\rm{in}}}{n_{\rm{out}}u^{r}_{\rm{out}}}E_{\rm{in}}
=-|\frac{n_{\rm{in}}u^{r}_{\rm{in}}}{n_{\rm{out}}u^{r}_{\rm{out}}}|E_{\rm{in}}\;, 
\end{equation}
where the subscript "in" and "out", respectively denotes the quantities for the inflow and the outflow. The above equation 
is obtained by considering the conservation of the outward energy flux near the separation point, provided that 
the magnetic field configuration does not change significantly there. In general, the ratio 
of the inflow and outflow particle flux, $|n_{\rm{in}}u^{r}_{\rm{in}}/n_{\rm{out}}u^{r}_{\rm{out}}|$, is
greater than unity because only a small portion of the plasmas will escape as an outflow.
As a result, Equations (35) and (36) give 
\begin{equation}
 \Gamma\geq-\frac{E_{\rm{in}}}{\mu}\;. 
\end{equation}
Hence, the inverse values of the total energy shown in Figures 4(d) and (h), as shown in Figure 6, 
give an estimation of the lower limit of the ergospheric jet Lorentz factor  for BHXBs and AGNs. 
The BHXB and AGN relativistic jet speeds inferred from the observations, which has the typical value $\Gamma>2$ for BHXBs 
(e.g., Table 1 of Fender et al. 2004) and $\Gamma\sim10$ for AGNs, can be  consistently explained by our model. 

\begin{figure}[h]
%\epsscale{.80}
\plotone{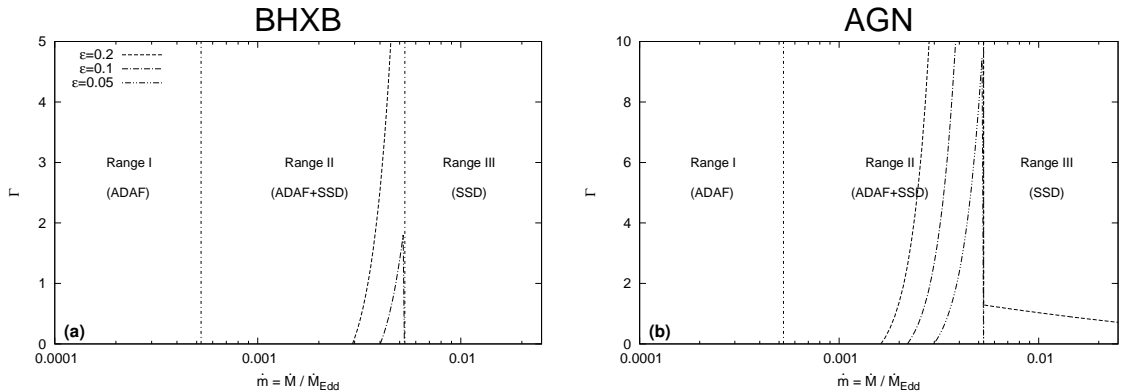}
\caption{Lower limits of the Lorentz factor of ergospheric jets, $\Gamma$, modeled for BHXBs (a) and 
AGNs (b), as a function of the accretion rate, $\dot{m}$. See text for more details. 
%When the relativistic jet is "turned on",
%Relativistic jet is turned on when the transition radius of the combined disk is close enough to the central black hole,
%which results in a magnetically dominated environment near the BH (see text), and the jet terminal speed rapidly increases
%with $\dot{m}$ before the entire disk become a thin disk.
\label{fig6}
}
\end{figure}

\subsection{Validity of the Model}
To investigate
the reason of the launching and quenching of BH relativistic jets
when $\dot{m}$ varies,
we solve for stationary negative energy inflow solutions 
(which corresponds to the formation of relativistic jets)
by adopting analytical solution of the ADAF and the SSD, 
assuming a spherical accretion geometry near
the horizon, and estimating the field
strength of the large-scale hole-threading magnetic field by the
parameter $\varepsilon$.
The major drawback of our approach is that we ignore the possible
disk instability when the SSD becomes radiation pressure dominated.
In addition, our axial-symmetry and stationary solution
miss the temporal and the non-axisymmetry properties
of the jet formation. Also,
the disk dynamo effect and the magnetic reconnection process can further modify the 
large-scale fields, although
their effects can be in general included in the parameter $\varepsilon$.

For our purpose, we ignore several aspects which we think are minor factors in
determining whether the 
jet is \textit{on} or \textit{off} for simplicity.
Several discussions on the assumptions and the robustness of the result 
are provided in the following.

We ignore the  Blandford-Znajek extraction  in the
force-free region near the pole of the horizon in a quasi-spherical accretion flow. 
However, such extraction is
expected to be a relatively minor factor to determine the 
jet launching.
The reason is that
most of the hole-threading field
lines  have considerable plasma loading on them, and that
the power for magnetic fields extracting the energy from a rotating BH
 is proportional to $\sin^{2}\theta$ \citep{bla77}.

The most important feature of a combined disk is that the significant 
surface density  difference of the SSD and the ADAF
at the transition radius.
Although here we describe the combined disk solution by patching
the analytical solutions of the SSD and the ADAF at
the transition radius,
the above feature
is expected to be true even if a more detailed disk solution is considered.

There are uncertainties  in determining the large-scale magnetic fields.
For example,
the parameter $\varepsilon$ can vary with the radius $r$, and 
the total flux of the large-scale disk-threading magnetic fields 
is not unclear. However, 
the field lines that
threads the disk are expected to be redistributed when the accretion disk types vary.
The disk-jet coupling features described by our model
can be qualitatively  preserved as long as the diffuse effect
is relatively unimportant than the advection effect for
the large-scale disk-threading field lines.
Especially,  a magnetically dominated magnetosphere near the horizon is still most likely
realized when the BH is surrounded by a combined disk,
provided that the large-scale magnetic field
advected inward from the outer SSD is large enough for the inner ADAF.
 Such requirement is
not difficult to be satisfied as the transition radius gradually decreases
to a small enough radius (Figures 4(a) and (e)).

The result presented in Figure 4  shows that a negative
energy flow may form when the BH is surrounded by a combined disk (Range II). 
The conclusion can actually still be valid when more detailed pictures
are considered. 
Below we discuss two possible concerns.
First, we consider whether the result would change
when numerical ADAF solutions, e.g., \citet{nar97} and \citet{gam98}, instead
of the self-similar solutions, Equations (B1)-(B3), are used. It is true that 
the self-similar
analytical solutions are not a good approximation near the
hole. For example, the causality requires a numerical solution of
$u^{\mathrm{ADAF}}$ be greater than Equation (B1)
near the BH, while the surface density $\sigma^{\mathrm{ADAF}}$ be
less than the self-similar solution, Equation (B3). Nevertheless, the
numerator in Equation (C2) will not change because of the mass conservation,
$\sigma u\sim H\rho u\varpropto\dot{M}$, provided that the disk height
$H$ remains unchanged. Since the magnetic field is predominantly
supported by the outer thin disk, $\bar{B}_{\mathrm{II}}$ is essentially
given by $B(R_{\mathrm{tr}})$. As a result, neither $\mu\eta$ nor
$\bar{B}_{\mathrm{II}}$ contains significant errors if we adopt the
self-similar solution of ADAF. Thus, the discussion given in this paper
is not vulnerable for details of the ADAF solutions.
Second, we discuss whether the estimation of the mass loading is
fair when $\bar{B}_{\mathrm{II}}=B(R_{\mathrm{tr}})+B(R_{\mathrm{inj}})\gg B(R_{\mathrm{inj}})$.
In this case, an MCAF 
is formed.  The highly
non-axisymmetry, irregular
MCAF will try to avoid the
strong field ranges by the magnetic Rayleigh-Taylor instability and result
in strong-field, low-density {}`magnetic islands' \citep{pun09}. As a result,
both $\rho^{\mathrm{ADAF}}$ and $u^{\mathrm{ADAF}}$ \textit{decrease}
in a magnetic island. Thus, even when $R_{\mathrm{m}}>R_{\mathrm{inj}}$,
the mass loading from the magnetic island located at $R_{\mathrm{inj}}$
is lower than the estimate with the self-similar solutions, Equations (B1) and
(B2). This means that the MHD inflow launched from the magnetic island
should be even more magnetically dominated and that the energy of
the inflow is still negative.

\section{Conclusions}
In this paper we focus on the fundamental question: 
Why relativistic jets from accreting BHs can be selectively launched
under a certain accretion condition?
Here we solve this problem by incorporating the plasma inertia effect,
 which have been neglected in previous studies, in conjunction 
with the electromagnetic effect 
on the extraction of BH energy.
At low accretion rate (e.g., $\dot{m}\lesssim0.1$), the accretion flow has
a quasi-spherical geometry near the BH event horizon.
As a result, whether the rotational energy of the BH can be effectively
extracted outward is determined by the inward rest-mass energy of the
accreting plasmas. When the outward electromagnetic energy flux dominates,
a  relativistic, ergospheric jet can be therefore launched;
on the contrary, when the inward rest-mass energy dominates, there is
no powerful relativistic jet.
By examining the MHD Penrose process for BHs surrounded
by different types of accretion disk at 
low accretion rates, we conclude that the
MHD inflow becomes magnetically dominated preferentially in a combined
disk, which consists of an outer SSD and an inner ADAF, to 
enable the launch of an ergospheric
jet. Such relativistic jets are quenched if a magnetic dominance
breaks down when the entire disk becomes an SSD. 
We also provide a universal paradigm of the coupling between the BH accretion disks and
their relativistic jets at low accretion rates. Several observed disk-jet connections 
are naturally explained. 
Future study on the relation between
the accretion geometry near the event horizon and the extraction of the BH energy
will be important to the understanding of the BH jet formation mechanisms.

\acknowledgments
We are grateful to Mike J. Cai
for discussions about the relativistic Bernoulli equation. We thank
C. F. Gammie, A. K. H. Kong, M. Nakamura, H. C. Spruit, R. E. Taam and 
anonymous referees
for helpful comments and suggestions. 
This work was supported by the National
Science Council (NSC) of Taiwan under the grant NSC
99-2112-M-007-017-MY3 and the Formosa Program between NSC and Consejo
Superior de Investigaciones Cientificas in Spain administered under
the grant NSC100-2923-M-007-001-MY3.

\appendix
\section{Parameterized the Large-Scale Magnetic Field Strength}

The large-scale field $B$ stored in a disk can be estimated from the point of view of
 \textit{energies}. The large-scale field energy $E_{\mathrm{F}}$
inside the volume enclosed by a radius $R$ should not exceeds the
gravitational binding energy $E_{\mathrm{G}}$ of the disk near $R$,
otherwise the field would escape by buoyancy. By considering $E_{\mathrm{F}}\sim4\pi R^{3}B^{2}/24$
and $\mid E_{\mathrm{G}}\mid\sim2\pi GM_{\mathrm{BH}}\sigma R$, we define a ratio
$\varepsilon^{2}$ $(\leq1)$ between this two quantities to obtain \[
B\sim\varepsilon\sqrt{12GM_{\mathrm{BH}}\sigma/R^{2}}\;,\]
where $G$ is the gravitational constant, $M_{\mathrm{BH}}$ is the mass of the central
BH, and $\sigma$ is the surface density of the disk. 

Alternatively, we can estimate the large-scale field $B$ stored in
a disk by a \textit{force} balance. If a disk accretion is supported
by a large-scale field at $R$, as described in \citet{nar03},
the large-scale field $B$ can be estimated by the force balance $GM_{\mathrm{BH}}\sigma/R^{2}\sim B_{r}B_{\theta}/2\pi$
to give $B\sim\sqrt{2\pi GM_{\mathrm{BH}}\sigma/R^{2}}$ where $B_{\theta}\sim B_{r}\sim B$
is assumed. Similarly, a parameter $\varepsilon^{2}(\leq1)$ can be
defined by the ratio between the gravity force and the field force;
we thus obtain \[
B\sim\varepsilon\sqrt{2\pi GM_{\mathrm{BH}}\sigma/R^{2}}\;.\]

Since the two estimations above are consistent in order of magnitude,
we simply parameterize $B$ in terms of $\varepsilon$, $M_{\mathrm{BH}}$, and
$\sigma/R^{2}$, such that 

\begin{equation}
B\equiv B(\varepsilon,R,\sigma(R))=\varepsilon\sqrt{2\pi GM_{\mathrm{BH}}\sigma/R^{2}}\;.\end{equation}

\section{Analytical Solution of the Accretion Disks}

With the scaled units, \[
M_{\mathrm{BH}}=m\mathrm{M_{\odot\;,}}\]

\[
R=\bar{r}R_{\mathrm{g}}\qquad R_{\mathrm{g}}\equiv\frac{2GM_{\mathrm{BH}}}{c^{2}}=(2.95\times10^{5}m)\:\mathrm{cm}\;,\]

\[
\dot{M}=\dot{m}\dot{M}_{\mathrm{Edd}}\qquad\dot{M}_{\mathrm{Edd}}=L_{\mathrm{Edd}}/0.1c^{2}=(1.39\times10^{18}m)\:\mathrm{g/s}=(2.2\times10^{-8}m)\:\mathrm{M_{\odot}/yr}\;,\]
we adopt the following radial velocity $u$, plasma density $\rho$,
and surface density $\sigma$ solutions of an ADAF and a thin disk. 
Note that what we
need is $\rho u$ (because $\mu\eta\equiv\rho u/B$) to solve the
BE, and that $\rho u$ can be more or less correctly evaluated even
in the innermost region by an ADAF self-similar solution without
invoking a numerical solution. We therefore
adopt ADAF self-similar solution \citep{nar95,nar98} and obtain

\begin{equation}
u^{\mathrm{ADAF}}=-1.1\times10^{10}(\alpha^{\mathrm{ADAF}})\bar{r}^{-\frac{1}{2}}\;\mathrm{cms^{-1}}\;,\end{equation}

\begin{equation}
\rho^{\mathrm{ADAF}}=1.31\times10^{-4}(\alpha^{\mathrm{ADAF}})^{-1}m^{-1}\dot{m}\bar{r}^{-\frac{3}{2}}\quad\mathrm{gcm^{-3}\;},\end{equation}

\begin{equation}
\sigma^{\mathrm{ADAF}}=\dot{M}/2\pi R(-u^{\mathrm{ADAF}})=68(\alpha^{\mathrm{ADAF}})^{-1}\dot{m}\bar{r}^{-\frac{1}{2}}\quad\mathrm{gcm^{-2}}\;,\end{equation}
while in the thin disk solutions \citep{sha73,kat08}, we obtain

\begin{equation}
u^{\mathrm{SSD}}=-4.3\times10^{6}(\alpha^{SSD})^{\frac{4}{5}}m^{-\frac{1}{5}}\dot{m}^{\frac{2}{5}}\bar{r}^{-\frac{2}{5}}f^{-\frac{3}{5}}\mathrm{cms^{-1}}\;,\end{equation}

\begin{equation}
\rho^{\mathrm{SSD}}=20\times(\alpha^{SSD})^{-\frac{7}{10}}m^{-\frac{7}{10}}\dot{m}^{\frac{2}{5}}\bar{r}^{-\frac{33}{20}}f^{\frac{2}{5}}\quad\mathrm{gcm^{-3}}\;,\end{equation}

\begin{equation}
\sigma^{\mathrm{SSD}}=1.7\times10^{5}(\alpha^{SSD})^{-\frac{4}{5}}m^{\frac{1}{5}}\dot{m}^{\frac{3}{5}}\bar{r}^{-\frac{3}{5}}f^{\frac{3}{5}}\quad\mathrm{gcm^{-2}}\;,\end{equation}
where

\[
f=[1-(\frac{\bar{r}_{\mathrm{ISCO}}}{\bar{r}})^{\frac{1}{2}}]\;,\]
and $\bar{r}_{\mathrm{ISCO}}$ is the radius of the innermost stable
circular orbit. Note the dimensionless coordinate, $\bar{r}$, is
related to $r$ by $\overline{r}=2r$. The parameter $\alpha$ describes
the viscosity in the $\alpha$-prescription \citep{sha73}. 
The superscripts
{}``ADAF'' and {}``SSD'' denote the Advection-dominated accretion flow and the Shakura-Sunyaev disk analytical
solutions, respectively. 
To elucidate
the general feature, we adopt typical values of the viscosity parameter
such that $\alpha^{\mathrm{SSD}}=0.01$ and $\alpha^{\mathrm{ADAF}}=0.1$.

\section{Solving the Relativistic Bernoulli Equation for the Location of the Alfven Point}

Once the field strength, field geometry, and the mass loading per
field line at different $\dot{m}$ are specified, we can calculate
all the coefficients in the cold BE, Equation (26), and determine the
location of the Alfven point, $R_{\mathrm{A}}$. 

Since Equation (26) is solved in the geometrized unit (in length unit
of $GM/c^{2}$), extra transformations of units from c.g.s. unit are
necessary, namely,

\begin{equation}
\mathbb{\mathcal{H}}(\mathrm{geom.})\backsimeq\bar{B}(\mathrm{c.g.s})\times M_{BH}G^{3/2}/c^{4}\;,\end{equation}
 \begin{equation}
\mu\eta(\mathrm{geom.})\backsimeq m_{\mathrm{p}}n\frac{u}{\bar{B}}(\mathrm{c.g.s})\times M_{BH}G^{3/2}/c^{3}=\rho\frac{u}{\bar{B}}(\mathrm{c.g.s})\times M_{BH}G^{3/2}/c^{3}\;.\end{equation}
These unit transformations can be done by two steps. First, we convert
the unit from the c.g.s into the geometrized unit (Table 3). Second,
we rescale the length unit in geometrized unit from {}``$\mathrm{cm}$''
into {}``$GM/c^{2}$'' (Table 4). 

Next, we find the location of the Alfven point $R_{\mathrm{A}}$ at
various accretion rate by the following steps:
\begin{enumerate}
\item \noindent Calculate the coefficients $\mathcal{A}$ and $\mathcal{B}$
in Equation (26).
\item \noindent Guess $R_{\mathrm{A}}$ to compute $E$ and $L$ by Equations
(29) and (30).
\item \noindent Using $E$ and $L$, calculate the coefficients $\mathcal{C}$
and $\mathcal{D}$ in Equation (26).
\item \noindent Using the calculated $\mathcal{A}$ , $\mathcal{B}$, $\mathcal{C}$,
$\mathcal{D}$, solve the cold BE equation(26).
\item \noindent Find $R_{\mathrm{A}}$ so that the fluid can smoothly pass
through the fast-magnetosonic point.
\end{enumerate}
A correct $R_{\mathrm{A}}$ will give a physical inflow solution that
passes through the fast point before reaching the horizon with a nonzero
speed (Figure 7). If $R_{A}$ is incorrect, the resulting inflow solution
becomes unphysical, being either sub-fast-magnetosonic (Figure 8),
or discontinuous (Figure 9).

\begin{table}[h]
%\begin{table}
\begin{centering}
%\begin{center}
\caption{Unit Conversion from c.g.s. into Geometrized Unit\label{table3}}
\begin{tabular}{>{\centering}m{3cm}>{\centering}m{3cm}>{\centering}m{4cm}>{\centering}m{5cm}}
%{ccc>{\centering}p{4cm}}
\hline 
\hline
Quantity & c.g.s Unit & Geometrized Unit & Dimension of the Geometrized Unit\tabularnewline

\hline 
Rest-mass energy & $\mu=m_{\mathrm{p}}c^{2}$ & $\mu(\mathrm{c.g.s})G/c^{4}=m_{\mathrm{p}}(\mathrm{c.g.s})G/c^{2}$ & {[}$\mathrm{cm}${]}\tabularnewline
Number density & $n$ & $n(\mathrm{c.g.s})$ & {[}$\mathrm{cm^{-3}}${]}\tabularnewline
Magnetic field & $B_{\mathrm{p}}$ & $B_{\mathrm{P}}(\mathrm{c.g.s})G^{\frac{1}{2}}/c^{2}$ & {[}$\mathrm{cm^{-1}}${]}\tabularnewline
Velocity & $u_{\mathrm{p}}$ & $u_{\mathrm{P}}(\mathrm{c.g.s})/c$ & {[}$1${]}\tabularnewline
Mass loading & $\eta=nu_{\mathrm{p}}/B_{\mathrm{p}}$ & $\eta(\mathrm{c.g.s})c/G^{\frac{1}{2}}$ & {[}$\mathrm{cm^{-2}}${]}\tabularnewline
\hline
\end{tabular}
\par\end{centering}

\centering{}%

\end{table}

\begin{table}[h]
%\begin{table}
\begin{centering}
\caption{ Converting the Length Unit from {}``$\mathrm{cm}$'' to
{}``$GM/c^{2}$''\label{table4}}
\begin{tabular}{>{\centering}m{3cm}>{\centering}m{3cm}>{\centering}m{5cm}>{\centering}m{4cm}}
%{ccc>{\centering}p{3.5cm}}
\hline 
\hline
Quantity & Geometrized Unit & Geometrized Unit Scaled by $GM/c^{2}$ & Dimension after Rescaled\tabularnewline

\hline 
Rest-mass energy & $m_{\mathrm{p}}(\mathrm{c.g.s})G/c^{2}$ & {[}$m_{\mathrm{p}}(\mathrm{c.g.s})G/c^{2}]/[(GM/c^{2})${]} & {[}$(GM/c^{2})${]}\tabularnewline
Number density & $n(\mathrm{c.g.s})$ & {[}$n(\mathrm{c.g.s})]/[(GM/c^{2})^{-3}${]} & {[}$(GM/c^{2})^{-3}${]}\tabularnewline
Magnetic field & $B_{\mathrm{P}}(\mathrm{c.g.s})G^{\frac{1}{2}}/c^{2}$ & {[}$B_{\mathrm{P}}(\mathrm{c.g.s})G^{\frac{1}{2}}/c^{2}]/[(GM/c^{2})^{-1}]$ & {[}$(GM/c^{2})^{-1}${]}\tabularnewline
Velocity & $u_{\mathrm{P}}(\mathrm{c.g.s})/c$ & $u_{\mathrm{P}}(\mathrm{c.g.s})/c$ & {[}$1${]}\tabularnewline
Mass loading & $\eta(\mathrm{c.g.s})G/c^{2}$ & {[}$\eta(\mathrm{c.g.s})G/c^{2}]/[(GM/c^{2})^{-2}${]} & {[}$(GM/c^{2})^{-2}${]}\tabularnewline
\hline
\end{tabular}
\par\end{centering}

\centering{}%
\end{table}

\begin{figure}[h]
\epsscale{.50}
\plotone{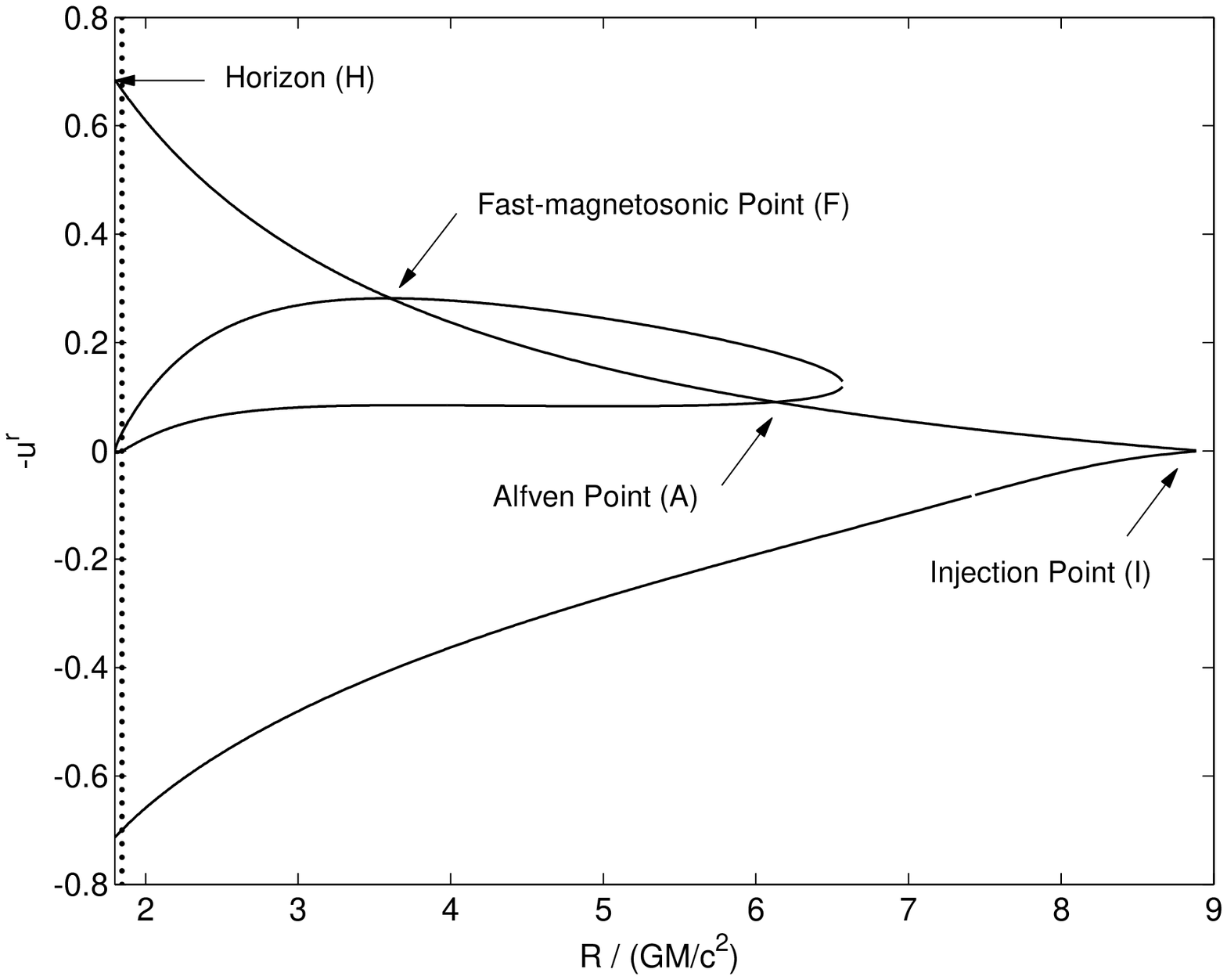}
\caption{Solution of the BE, Equation (26), when $M/M_{\odot}=10$,
$\varepsilon=0.2$ , $\dot{m}=0.001$, and $R_{\mathrm{A}}/(GM/c^{2})=6.128$.
The BE is a fourth-order algebraic equation and it has four roots
at a fixed $R$ in general. However, only the real solutions are shown
here. This physical solution (I$\rightarrow$A$\rightarrow$F$\rightarrow$H)
shows that the fluid is injected at the injection point (I) and passes
through the Alfven (A) and the fast-magnetosonic (F) points, finally
reaches the black hole (H) with a super-fast-magnetosonic speed. The
vertical dotted line indicates the location of the inner light surface.
We are not interested in the outward solution ($-u^{r}<0$), because
it represents an MHD flow escaping from the BH.\label{fig7}}
\end{figure}

\begin{figure}[h]
\epsscale{.50}
\plotone{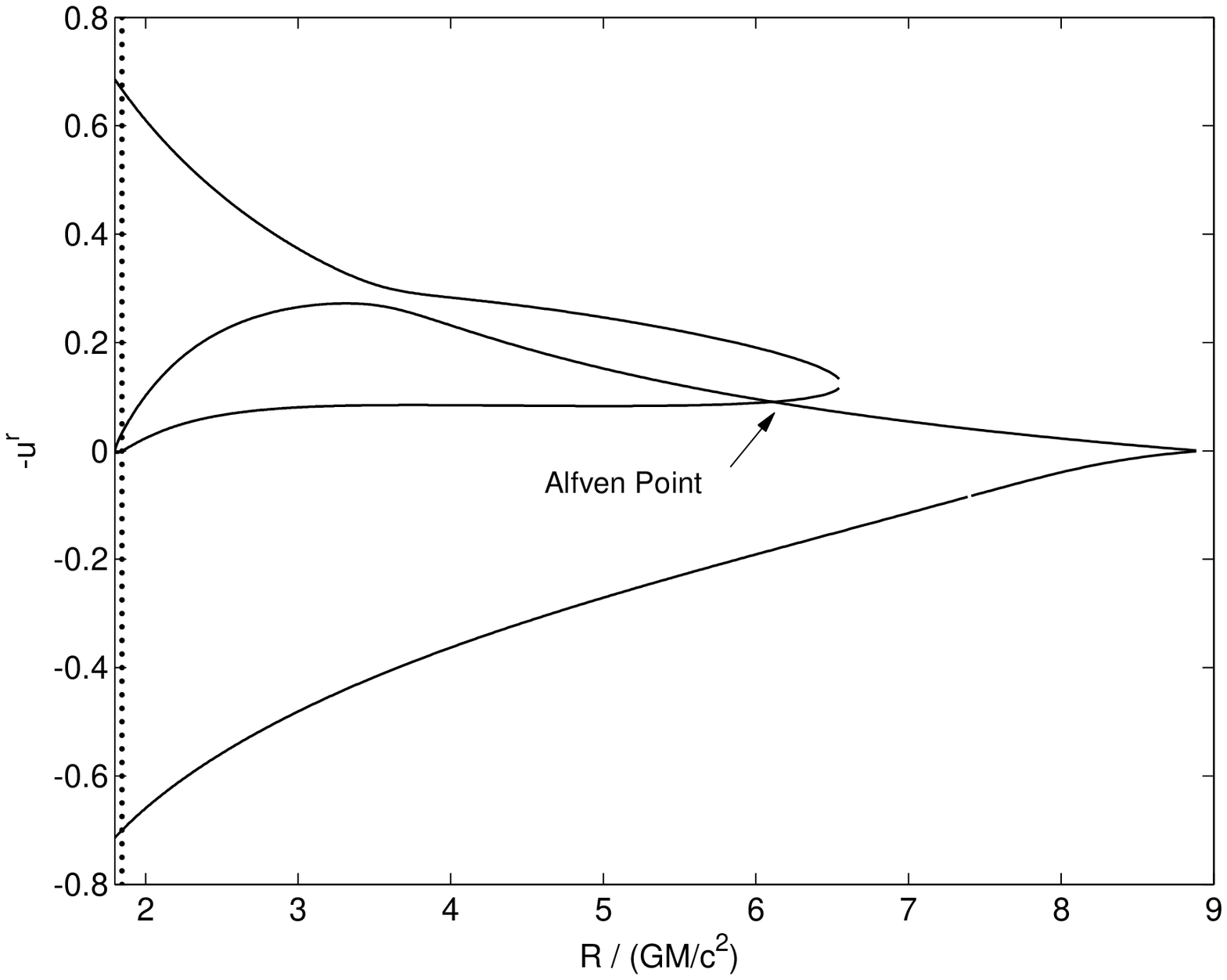}
\caption{Same as Figure 7, but for $R_{\mathrm{A}}/(GM/c^{2})=6.109$.
The solution is unphysical because the trajectory starting from the
injection point reaches the horizon with a vanishing inward velocity.\label{fig8}}
\end{figure}

\begin{figure}[h]
\epsscale{.50}
\plotone{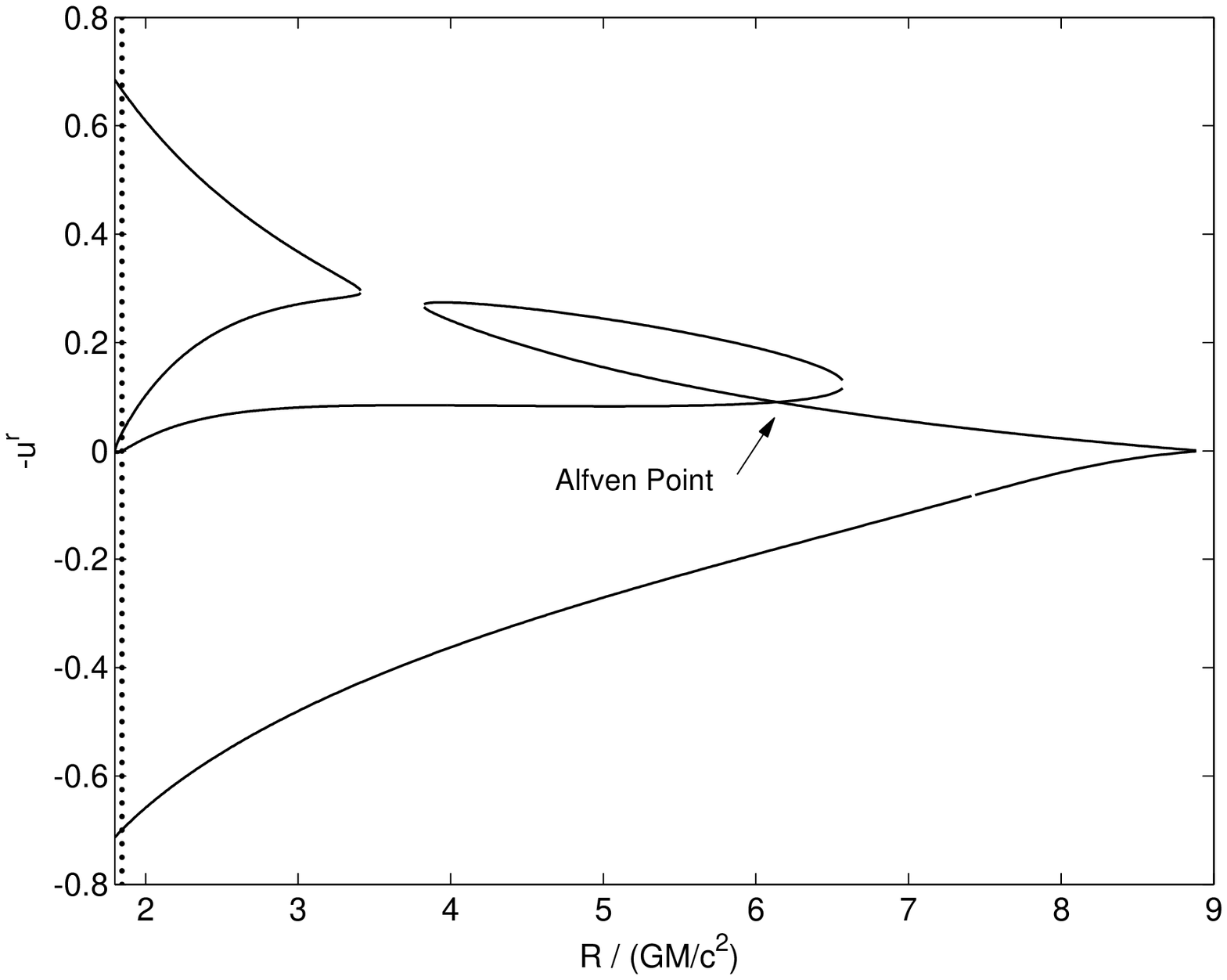}
\caption{Same as Figure 7, but for $R_{\mathrm{A}}/(GM/c^{2})=6.138$.
The solution that passes the Alfven point is unphysical, because the
flow cannot reach the horizon.\label{fig9}}
\end{figure}

\end{document}